\documentclass{article}
\usepackage[utf8]{inputenc}

\usepackage{color}
\usepackage{amsmath}
\usepackage{tabularx}
\usepackage{amsfonts}
\usepackage{graphicx}
\usepackage[dvipsnames]{xcolor}
\definecolor{refs}{RGB}{245,156,74}
\usepackage[colorlinks=cyan,hyperfootnotes=true,citecolor=cyan]{hyperref}
\usepackage{footnote}
\usepackage{footmisc}
\makesavenoteenv{tabular}
\makesavenoteenv{table}

\renewcommand{\thefootnote}{\alph{footnote}}

\newcommand{\astfootnote}[1]{%
	\let\oldthefootnote=\thefootnote%
	\setcounter{footnote}{0}%
	\renewcommand{\thefootnote}{\fnsymbol{footnote}}%
	\footnote{#1}%
	\let\thefootnote=\oldthefootnote%
}

\renewcommand{\thefootnote}{\alph{footnote}}

\newcommand{\dagfootnote}[1]{%
	\let\oldthefootnote=\thefootnote%
	\setcounter{footnote}{1}%
	\renewcommand{\thefootnote}{\fnsymbol{footnote}}%
	\footnote{#1}%
	\let\thefootnote=\oldthefootnote%
}

\renewcommand{\thefootnote}{\alph{footnote}}

\newcommand{\thirdfootnote}[1]{%
	\let\oldthefootnote=\thefootnote%
	\setcounter{footnote}{2}%
	\renewcommand{\thefootnote}{\fnsymbol{footnote}}%
	\footnote{#1}%
	\let\thefootnote=\oldthefootnote%
}

\newcommand{\be}{\begin{equation}}
\newcommand{\ee}{\end{equation}}
\newcommand{\bea}{\begin{eqnarray}}
\newcommand{\eea}{\end{eqnarray}}

\newcommand{\tetrad}{\theta}
\newcommand{\cotetrad}{e}

\newcommand{\spinconnection}{\omega}

\newcommand{\Lorentz}{\Lambda}

\newcommand{\lapse}{\alpha}
\newcommand{\shift}{\beta}
\newcommand{\inducedmetric}{\gamma}
\newcommand{\normalvector}{\xi}
\newcommand{\momenta}{\pi}

\title{The  Hamilton equations in $f(T)$ teleparallel gravity and in New General Relativity}

\author{Francesco Bajardi${}^{1,2}$\astfootnote{\href{mailto:francesco.bajardi@unina.it}{f.bajardi@ssmeridionale.it}},  Daniel Blixt$^{1}$\dagfootnote{\href{mailto:d.blixt@ssmeridionale.it}{d.blixt@ssmeridionale.it}} \ and Salvatore Capozziello${}^{1,2,3}$  \small \thirdfootnote{\href{mailto:capozziello@na.infn.it}{capozziello@na.infn.it}}, 
	\\
	{\small $^1$ {\it Scuola Superiore Meridionale,}}
	{\small {\it Largo S. Marcellino 10, I-80138 Napoli, Italy}}\\
	{\small $^2$ {\it Istituto Nazionale di Fisica Nucleare, Sezione di Napoli, Via Cintia 80126 Napoli, Italy.}}\\
{\small $^3$ {\it Dipartimento di Fisica "E. Pancini",Universit\`a degli Studi di Napoli "Federico II"}}\\
		{\small {\it  Complesso Universitario di Monte S. Angelo, via Cinthia, Ed. N, 80126 Napoli, Italy}}}

\begin{document}
	
	\maketitle
	
	\begin{abstract}
		We derive the Hamiltonian function for extended teleparallel  theories of gravity in their covariant formulation. In particular, we present the Hamiltonian for $f(T)$ gravity and New General Relativity. From this, we obtain the related Hamilton equations, which are presented both in  covariant formulation and  Weitzenböck gauge. In this framework,    teleparallel equivalent to General Relativity, its $f(T)$ extension and New General Relativity can be compared. We find that $f(T)$ and New General Relativity consistently reduce to the Teleparallel Equivalent to General Relativity, while significant differences appear comparing the Hamilton equations of $f(T)$ with $f(R)$ gravity.
	\end{abstract}
	
	\section{Introduction}
	
	After more than one century from its formulation, General Relativity (GR) has been confirmed by numerous experiments and observations, and it remains an essential part of our understanding of gravity and the Universe. However, though it is a highly successful theory of gravity - the best accepted thus far - it manifests some shortcomings and limitations \cite{Will:2014kxa}. For instance, it is incompatible with Quantum Mechanics \cite{Goroff:1985th}, which governs the dynamics at very small scales; it cannot  explain phenomena such as dark matter and dark energy, which are believed to make up the majority of the Universe content \cite{Frieman:2008sn, Mannheim:2005bfa}; it predicts the existence of singularities, where standard laws of physics break down \cite{Barack:2018yly}; it does not provide a self-consistent theory of Quantum Gravity, which would merge GR and Quantum Mechanics into a single, coherent picture \cite{DeWitt:1967yk}. For these reasons, alternative theories of gravity have been proposed to address specific issues with GR, such as the existence of dark matter and dark energy, the formation of structure in the Universe, and the behavior of gravity at early times \cite{Copeland:2006wr,Bamba:2012cp,Capozziello:2002rd,Capozziello:2003tk, Sanders:2002pf, Famaey:2011kh, Nojiri:2010wj, Capozziello:2019cav}. Some of the most well-known modified theories of gravity include \emph{e.g.} Modified Newtonian Dynamics (MOND) \cite{Bekenstein:2004ne}, which proposes a modification of Newton's law of gravitation to account for observed discrepancies in the motion of celestial bodies; Brans-Dicke theory \cite{Brans:1961sx}, which replaces, in agreement with the Mach principle,  the Newtonian constant with a scalar field and allows for the possibility of variations in the strength of gravity over time and space; $f(R)$ gravity \cite{Capozziello:2002rd,DeFelice:2010aj,Capozziello:2011et, Starobinsky:2007hu, Nojiri:2006gh,Nojiri:2017ncd}, which extends the Einstein-Hilbert action, linear in the Ricci scalar $R$,  to a generic function of such a scalar invariant; scalar-tensor theories \cite{Clifton:2011jh, Bajardi:2020xfj, Damour:1992we, Gleyzes:2014dya}, which generalize GR by including additional scalar fields and can influence the gravitational force; Gauss--Bonnet gravity, including into the gravitational action the Gauss-Bonnet topological surface \cite{Bajardi:2022tzn, Bajardi:2020mdp, Nojiri:2005jg, Cognola:2006eg, Bajardi:2024efo}; higher-dimensional theories, which aim to fix small-scale issues by increasing the number of dimensions \cite{Bajardi:2021hya, Gibbons:1987ps, Schwarz:1982jn, Emparan:2008eg}. Most of them lead to modifications (and extensions)  of the Newtonian potential \cite{Capozziello:2020dvd,Capozziello:2021goa}. However, they are still being developed and tested, and it is not yet clear which, if any, will become the dominant theory of gravity in the future. By relaxing the assumption of symmetric connection with respect to the lowest indexes, it is possible to introduce  torsion in the spacetime, dealing with both curvature and torsion. This formalism, considered \emph{e.g.} in \cite{Poplawski:2010kb, Cabral:2019gzh} is called the \emph{Einstein-Cartan Formalism}. In some cases, this leads to the breaking of the Equivalence Principle \cite{Arcos:2004tzt} and allows to describe gravity at small scales \cite{Casadio:2021zai, Krssak:2015lba}. In particular, imposing the spacetime to be governed only by torsion instead of curvature, it is possible to develop a self-consistent theory of gravity, whose dynamics is exactly the same as GR. This theory is called \emph{Teleparallel Equivalent to General Relativity} (TEGR) \cite{Maluf:2013gaa, Bahamonde:2015zma}. The latter has been deeply studied in the last years and has been the subject of numerous studies and investigations \cite{Xu:2012jf, Krssak:2018ywd, Obukhov:2002tm, Geng:2011ka, Bahamonde:2021gfp}. This approach represents a theoretical framework to describe gravity which is based on the concept of parallelism instead of curvature: here gravitational dynamics is described as the result of torsion in the spacetime fabric. In teleparallelism, the gravitational potentials are  a set of tetrad fields (also known as ``vierbeins''), which form a basis for describing the geometry of spacetime. These tetrad fields are used to define a torsion tensor, which acts as the source of gravity  and represents the anti-symmetric contribution of the Christoffel connection. The gravitational action is then made of the  "Torsion Scalar", defined as a particular contraction of the Torsion Tensor. However, being completely equivalent to GR at the level of  field equations, TEGR cannot address issues and limitations provided by the Einstein theory at properly large scales. For this reason, in analogy with $f(R)$ gravity in the metric formalism, the Lagrangian density of TEGR can be modified and extended  in several way \cite{Bajardi:2021tul}, \emph{e.g.} by an arbitrary function of the torsion scalar, giving rise to the so called $f(T)$ gravity \cite{Cai:2015emx, Li:2010cg}. The latter has been proposed as a way to address shortcomings in the late-time, such as the accelerated expansion of the Universe \cite{Ferraro:2006jd, Wu:2010xk}, providing new types of solutions and the existence of alternative models. However, so far, it is not clear whether $f(T)$ can provide a better explanation of the observed behavior of gravity than GR, and more research is needed to determine its viability as a self-consistent  theory of gravity. See Refs.~\cite{Capozziello:2022zzh, Aviles:2013nga, Capozziello:2018hly} for further details on foundations and applications of teleparallel theories. 
	
	Another extension of TEGR called \emph{New General Relativity} (NGR) was proposed in \cite{Hayashi:1979qx}. Unlike $f(T)$, NGR is not a nonlinear extension, but instead the modification consists of adding torsion contractions at the same order of derivatives as in TEGR. All NGR theories, except one, has been disregarded in the literature due to the claims in Ref. \cite{Kuhfuss:1986rb}, for which only a particular case is ghost-free. This argument  was recently found to be incorrect \cite{Golovnev:2023jnc,Bahamonde:2024zkb,Tomonari:2024lpv,Tomonari:2024ybs}. The particular theory is motivated by being the only ghost-free extension of TEGR under these assumptions and it was also found that the PPN-parameters coincides with those of GR. Thus, it is believed that the theory is consistent with Solar System tests (which is also true for $f(T)$) \cite{Iorio:2012cm}.  However, recent findings proved that $f(T)$ and NGR contain  strongly coupled field \cite{Blixt:2020ekl, BeltranJimenez:2020fvy}, indicating that we need to carefully investigate the validity of observational predictions such as the PPN-parameters. This can be motivated by the presence of screening mechanisms \cite{Bahamonde:2024zkb} which could imply a recovery of the GR-limit at Solar System scales. As $f(T)$ gravity extends TEGR by introducing into the gravitational action a function of $T$, NGR aims to extend the definition of the torsion scalar by means of a more general quantity, $T_{NGR}$, to introduce into the gravitational action. Clearly, in this way, TEGR accounts for a particular sub-case of NGR. 
    
    Here,  we are going to consider a function of $T_{NGR}$ and find the Hamilton equations by means of a 3+1 decomposition. The latter is a mathematical technique used to describe the spacetime geometry and evolution of the Universe. The key idea is to use a spacetime foliation, which is a way of dividing spacetime into a series of space-like hypersurfaces that are labeled by a time coordinate. This allows one to describe  geometry and evolution of spacetime in terms of quantities that are defined on these hypersurfaces, such as the 3-metric and the extrinsic curvature. The 3+1 decomposition is widely used in numerical relativity and has proven to be a powerful tool for studying a wide range of problems in GR (and beyond), from the evolution of black hole spacetimes to the dynamics of the early Universe \cite{Elbistan:2022plu, Arnowitt:1962hi,Capozziello:2012hm}. 
    
    In this paper, we want to critically discuss the Hamilton equations in extended teleparallel theories as $f(T)$ gravity and compare them with analogous in NGR and then $f(R)$ gravity  to put in evidence differences among the  theories.  
	
	The manuscript is organized as follows: in Sec. \ref{sec:tel} we briefly overview the main features of TEGR, NGR (and their extensions) as well as the 3+1 decomposition of the metric. In Sec. \ref{sec:Ham} we obtain the Hamiltonian for $f(T_{NGR})$ gravity.  The Hamilton equations for the limiting cases of $f(T)$ gravity and NGR are presented in Secs. \ref{sec:HeqfT} and \ref{sec:HeqNGR}, respectively. The approach is developed in both the covariant formulation and the Weitzenböck gauge. Finally, in Sec. \ref{sec:conc}, we draw conclusions  with a final discussion and future perspectives. The Hamilton equations of  $f(T_\mathrm{NGR})$ gravity  are presented in detail in   App \ref{sec:Heqscov}. 
	
	\section{Teleparallel gravity and tetrad 3+1 decomposition}
	\label{sec:tel}
	Let us present now an introduction to teleparallel theories and the 3+1 decomposition, which will be needed in the derivation of the Hamiltonian function for  teleparallel models. We adopt the following conventions. Greek indexes denote coordinate indexes in four dimensions (running from 0 to 3), while lower case Latin indexes denote spatial coordinates (running from 1 to 3). Lorentz indexes running from 0 to 3 are denoted by capital Latin indexes. We use the mostly positive sign convention $\mathrm{diag}\eta_{AB}=(-1,1,1,1)$. To shorten certain expressions, indexes are sometimes placed in a non-canonical position. To obtain the canonical positions, the indexes have to be raised or lowered with the metric corresponding to the manifold where the index is defined (an example is presented in Eq. \eqref{eq:shorthand}). Overall, the notation coincides with that of Ref. \cite{Pati:2022nwi}, to simplify the comparison with the results for TEGR. This section introduces TEGR and extended teleparallel theories in Sec. \ref{sec:TEGR}. In Sec. \ref{sec:3+1}, the 3+1 decomposition is introduced (as well as the notation related to it). 
	
	\subsection{Teleparallel equivalent to General Relativity and its extensions}
	\label{sec:TEGR}
As previously mentioned, GR is the result of different assumptions ranging from the functional form of the action, the Equivalence Principle, up to the symmetry properties of the affine connection. In particular, the latter is supposed to be symmetric with respect to the lowest indexes, with the consequence that the spacetime turns out to be described only by curvature. In this way, the connection cannot be disentangled by the metric tensor and the action can be uniquely determined once assigning the line-element form. Moreover, if one also breaks the validity of the metricity condition, namely imposing the covariant derivative of the metric to vanish, it is possible to introduce the most general connection as follows \cite{Shabani:2017owe, Sotiriou:2009xt}:
\begin{equation}
\Gamma^\alpha_{\,\, \, \mu \nu} = \hat{\Gamma}^\alpha_{\,\,\, \mu \nu} + \frac{1}{2}g^{\alpha\lambda}\bigl(T_{\mu\lambda\nu}+T_{\nu\lambda\mu}+T_{\lambda\mu\nu}\bigr) + \frac{1}{2}g^{\alpha\lambda}\bigl(-Q_{\mu \nu \lambda}-Q_{\nu \mu \lambda} + Q_{\lambda\mu\nu}\bigr),
\label{genconn}
\end{equation} 
where $\hat{\Gamma}^\alpha_{\,\,\, \mu \nu}$ is the Levi-Civita connection, $T_{\mu\lambda\nu}$ is the  \emph{Torsion Tensor}, defined as $T^{\rho}_{\,\,\,\mu \nu} = 2 \Gamma^\rho_{\,\,\,[ \mu \nu ]}$, and $Q_{\mu \nu \lambda}$ is the \emph{Non-Metricity Tensor}, namely $Q_{\mu \nu \lambda} = \nabla_{\mu} g_{\nu \lambda}$, with $\nabla$ being the covariant derivative. The introduction of torsion implies that when a vector is parallel transported around a closed path, its final position will be shifted with respect to the initial one. On the other hand, non-metricity implies a spacetime in which the norm of a vector changes while parallel transported along a closed path, with the consequence that the manifold isometry is violated. 

	In the Einstein-Hilbert formulation, it is assumed that both torsion and non-metricity vanish. Under the metric teleparallel condition, where both the non-metricity and the Riemann tensor in a metric-affine geometry vanish, the Einstein-Hilbert action can be re-expressed in terms of torsion instead. Specifically, by defining the superpotential $S^{\rho \mu \nu}$ and the contortion tensor $K^{\rho}_{\,\,\,\,\mu\nu}$ as
 \begin{equation}
     \begin{split}
        & S^{\rho \mu \nu} \equiv K^{\mu \nu \rho}-g^{\rho \nu} T_{\,\,\,\,\,\,\, \sigma}^{\sigma \mu}+g^{\rho \mu} T_{\,\,\,\,\,\,\, \sigma}^{\sigma \nu}\,,
        \\
        & K^{\rho}_{\,\,\,\,\mu\nu}\equiv\frac{1}{2}g^{\rho\lambda}\bigl(T_{\mu\lambda\nu}+T_{\nu\lambda\mu}+T_{\lambda\mu\nu}\bigr)%
                                      =-K^{\rho}_{\,\,\,\,\nu\mu} \,,
     \end{split}
 \end{equation}
 respectively, one can define the  \emph{torsion scalar} as
  \begin{equation}
    T \equiv T^{\rho \mu \nu} S_{\rho \mu \nu}  \,.
    \label{torsiondef}
 \end{equation}
 By means of these definitions, it is straightforward to verify that the torsion and the curvature scalars only differ for a boundary term \cite{Capozziello:2017xla}. However, not providing contributions to the field equations, the latter is usually neglected. The formulation in which the boundary term is dropped is generally referred as TEGR, whose action reads:
	\begin{align}
		\label{eq:TEGR}
		S_\mathrm{TEGR}=-\frac{1}{2\kappa}\int \mathrm{d}^4 x   \, \sqrt{-g} \, T+S_\mathrm{M},
	\end{align}
		where $S_M$ is the matter action and $\kappa = 1/(8 \pi G)$, with $G$ being the Newton constant. As mentioned above, the TEGR action differs from the GR one only for a total divergence which, thus, does not contribute to the equations of motion.
  
  Interestingly, using the tetrad formulation, TEGR can be recast as a gauge theory with respect to the translation group in the local tangent spacetime. In doing so, the standard definition of tetrad fields, namely $\theta^A_\mu = \partial_\mu x^A$, must be generalized including both spin and linear connection:
  \begin{equation}
      \nabla_\mu x^A = \partial_\mu x^A + \omega^A_{\,\,\, B \mu} x^B - \Gamma^\alpha_{\,\,\, \mu \alpha} x^A,
  \end{equation}
where $\omega^A_{\,\,\, B \mu}$ is the spin connection. Using the tetrad postulate, according to which the covariant derivative of tetrad fields must vanish, it is possible to express the connection with the following relation
 \begin{equation}
     \Gamma^\alpha_{\,\,\, \mu \nu} = \theta^\alpha_A \mathcal{D}_\mu \theta^A_\nu ,
     \label{connectiontg}
 \end{equation}
 with $\mathcal{D}_\mu$ being the Lorentz covariant derivative: $\mathcal{D}_\mu x^A = \partial_\mu x^A + \omega^A_{\,\,\, B \mu} x^B $. Choosing the reference frame in which the spin connection vanishes, referred to as \emph{Weitzenböck gauge}, Eq. \eqref{connectiontg} becomes
 \begin{equation}
     \Gamma^\alpha_{\,\,\, \mu \nu} = \theta^\alpha_A \partial_\mu \theta^A_\nu .
 \end{equation}
 Denoting with $\theta$ the determinant of tetrad fields, the TEGR action \eqref{eq:TEGR} can be equivalently written as:
\begin{equation}
    S_{TEGR}=-\frac{1}{2\kappa}\int \mathrm{d}^4 x \tetrad \, T+S_\mathrm{M},
\end{equation}
and leads to the following field equations (in vacuum):
\begin{equation}
\frac{4}{\theta} \partial_{\mu}\left(\theta S_{A}^{\,\,\, \mu \beta}\right)-4 T_{\,\,\, \mu A}^{\sigma} S_{\sigma}^{\,\,\, \beta \mu}-T\, \theta_{A}^{\beta} = 0.
\label{teleparallel field equations}
\end{equation}
From Eq. \eqref{torsiondef}, one can straightforwardly notice the relation occurring between the Ricci and the torsion scalar, that is $R=-T+2\tetrad \overset{\circ}{\nabla}{}_\mu T^\mu$, with $\overset{\circ}{\nabla}{}_\mu$ being the covariant derivative expressed in terms of the Levi Civita connection and $T^\mu$ being a rank-1 tensor defined as
 \begin{equation}
    T^\mu =T_{\,\,\,\,\,\,\,\sigma}^{\sigma \mu}.
 \end{equation}

For our purpose, it is also worth noticing that the torsion scalar can be also expressed in terms of the torsion tensor as
 \begin{equation}
     T= -\frac{1}{4} T_{\mu\nu\rho}T^{\mu\nu\rho} -\frac{1}{2} T_{\mu\nu\rho}T^{\rho\nu\mu}+ T_\mu T^\mu. 
     \label{torsionscalar2}
 \end{equation}
 Teleparallel theories of gravity are often considered as modified theories of gravity since they naturally allow for theories with both symmetric and antisymmetric field equations (in contrast to GR and most of its modifications consisting of only symmetric field equations). Nonetheless, being equivalent to Einstein's theory at the level of equations, TEGR suffers the same shortcomings exhibited by GR at large scales. For this reason, in analogy with modifications of GR extending the gravitational action, several alternatives to standard TEGR have been considered. The most well-known modified teleparallel theories are $f(T)$ gravity and NGR. Inspired by $f(R)$ gravity, $f(T)$ gravity is given by the action
	\begin{align}
		\label{eq:fT}
		S_{f(T)}=-\frac{1}{2\kappa}\int \mathrm{d}^4 x \tetrad f(T)+S_\mathrm{M}.
	\end{align}
	Even though the TEGR and the Einstein-Hilbert actions give rise to the same field equations, the same is not true for $f(T)$ and $f(R)$. This is due to the fact that the boundary term provides a nontrivial contribution to the field equations, when the function $f$ acts on it 
    \cite{Capozziello:2023vne}.
    
	Another approach to modified teleparallel gravity is letting the coefficients appearing in the torsion scalar \eqref{torsionscalar2} to be arbitrary, by introducing a new torsion scalar, namely
	\begin{align}
	T_\mathrm{NGR}=c_1 T_{\mu\nu\rho}T^{\mu\nu\rho}+c_2 T_{\mu\nu\rho}T^{\rho\nu\mu}+c_3 T_\mu T^\mu. 
 \label{TNGR}
	\end{align}
	In this way, the TEGR action can be generalized as
		\begin{align}
	\label{eq:NGR}
	S_\mathrm{NGR}=-\frac{1}{2\kappa}\int \mathrm{d}^4 x \, \tetrad  \, T_\mathrm{NGR}+S_\mathrm{M}.
	\end{align}
	Furthermore, it was noted in \cite{Ortin:2015hya} that the condition $2c_1+c_2+c_3=0$ is required for this theory to be ghost-free. Note that, in order to have a propagating spin-2 field (as required in gravitational theories), with the correct gravitation strength, we further require that $c_3=1$. These two conditions leave us with a one-parameter ghost-free theory different from TEGR, generally called "the one-parameter theory of consistent NGR" \cite{Cheng:1988zg,BeltranJimenez:2019nns}.

	In \cite{Blixt:2020ekl}, the Hamiltonian for $f(T_\mathrm{NGR})$-gravity was derived in the Weitzenböck gauge, though the theory itself is not theoretically motivated. Nevertheless, it is the easiest that one can construct which reproduces both the most popular teleparallel theories (i.e. $f(T)$ and NGR). The action formulation of the theory is given by 
	\begin{align}
	S_{f(T_\mathrm{NGR})}=\frac{1}{2\kappa} \int \mathrm{d}^4x \, \theta f(T_\mathrm{NGR})+S_M,
	\end{align}
	or equivalently, in the Einstein frame
	\begin{align}
	S_{f(T_\mathrm{NGR})}=\frac{1}{2\kappa} \int \mathrm{d}^4x \, \theta \left(\phi T_\mathrm{NGR}-V(\phi)\right)+S_M\,,
	\end{align}
    where the scalar field $\phi$ represents the further degrees of freedom related to NGR with respect to TEGR.
    
	In Sec. \ref{sec:Ham} we will present the covariant Hamiltonian for this theory for the first time and App. \ref{sec:Heqscov} presents its Hamilton's equations. 
	
	\subsection{The 3+1 decomposition}
	\label{sec:3+1}
The 3+1 split\footnote{Here we follow the notation adopted in Ref. \cite{Pati:2022nwi}}  consists of three dimensional hypersurfaces of constant time slices $\Sigma_t$ and a normal vector $\normalvector^\mu$ orthogonal to $\Sigma_t$, which satisfies the condition $\normalvector_\mu \normalvector^\mu=-1$ \cite{Capozziello:2021pcg}. As pointed out  in Ref. \cite{Blixt:2020ekl}, this split is made for the spacetime indices only and not for the Lorentz indices. The hypersurfaces $\Sigma_t$ constitute a manifold with spatial indices $i,j,k,...$, equipped with the induced metric $\inducedmetric_{ij}$. According to this split, the tetrads become
	\begin{align}
	\tetrad^A{}_0 =\lapse \normalvector^A+\shift^i \tetrad^A{}_i,
	\end{align}
	where $\lapse$ is the lapse function, $\beta^i$ is the shift vector and 
	\begin{align}
	\normalvector^A =-\frac{1}{6}\epsilon^A{}_{BCD}\tetrad^B{}_i \tetrad^C{}_j \tetrad^D{}_k \epsilon^{ijk},
	\end{align}
	which satisfies the correct normalization property and is orthogonal to the spatial tetrads:
	\begin{align}
	\normalvector_A \tetrad^A{}_i=0.
	\end{align}
	The spatial tetrad, in turn, corresponds to the tetrad of the induced three-dimensional metric, that is:
	\begin{align}
	\inducedmetric_{ij}=\tetrad^A{}_i \tetrad^B{}_j \eta^{AB}.
	\end{align}
	It is also useful to define the cotetrad $\cotetrad_A{}^B$ in terms of the spatial tetrad, the shift vector, the lapse function and the spatial tetrad, as satisfying the identity
	\begin{align}
	\cotetrad_A{}^0=-\frac{1}{\lapse}\normalvector_A, \indent \cotetrad_A{}^i=\tetrad_A{}^i+\normalvector_A \frac{\shift^i}{\lapse}.
	\end{align}
	Here we are adopting the following shorthand notation
	\begin{align}
 \label{eq:shorthand}
	\tetrad_A{}^i=\tetrad^B{}_j \eta_{AB}\inducedmetric^{ij},
	\end{align}
	which will be used for brevity throughout the paper and where the indexes are placed at non-canonical positions. Using all of the above identities, it is straightforward to show that the well-known Arnowitt-Deser-Misner (ADM) decomposition of the metric (and its inverse) can be easily recovered:
	\begin{align}
	g_{\mu\nu}=\begin{bmatrix}
	-\lapse^2 +\shift^i \shift^j\inducedmetric_{ij} & \shift_i \\
	\shift_i & \inducedmetric_{ij}
	\end{bmatrix}, \indent g^{\mu\nu}=\begin{bmatrix}
	- \displaystyle \frac{1}{\lapse^2}   &  \displaystyle 
 \frac{\shift^i}{\lapse^2} \\
	 \displaystyle  \frac{\shift^i}{\lapse^2}  & \inducedmetric^{ij}- \displaystyle  \frac{\shift^i \shift^j}{\alpha^2}
	\end{bmatrix}.
	\end{align}
	After performing the aforementioned 3+1 split, the Lagrangian will only depend on the canonical variables $(\lapse,\shift^i, \tetrad^A{}_i,\Lambda_A{}^B, \phi)$ and their velocities (or functions of them, as for example the normal vector), with $\Lambda_A{}^B$ being Lorentz matrices. The application of such 3+1 split to the Lagrangian density results in
	\begin{align}
	\begin{split}
	\mathcal{L}_{f(T_\mathrm{NGR})}&=\frac{\sqrt{\inducedmetric}}{2\alpha}M^{i \ j}_{\ A \ B}T^A{}_{0i}T^B{}_{0j}-\frac{\sqrt{\inducedmetric}}{\lapse}T^A{}_{0i}T^B{}_{kl}\left[M^{i \ l}_{\ A \ B}\shift^k+\frac{\lapse\phi}{\kappa}\inducedmetric^{il}\left(c_2 \normalvector_B \tetrad_A{}^k+c_3 \normalvector_A \tetrad_B{}^k \right) \right]\\
	&+\mathcal{H}_S,
	\end{split}
 \label{LFt}
	\end{align}
	where $c_i$ are constant coefficients and
	\begin{align}
	M^{i \ j}_{\ A \ B}=-\frac{\phi}{\kappa}\left(2c_1 \inducedmetric^{ij}\eta_{AB}-(c_2+c_3)\normalvector_A \normalvector_B \inducedmetric^{ij}+c_2 \tetrad_A{}^j \tetrad_B{}^i +c_3 \tetrad_A{}^i \tetrad_B{}^j \right),
	\end{align}
	which, except for the overall factor $\displaystyle \frac{\sqrt{\inducedmetric}}{2\lapse}$, is the Hessian matrix. In Eq. \eqref{LFt}, the term $\mathcal{H}_S$ is defined as:
	\begin{align}
	\begin{split}
	\mathcal{H}_S=\frac{\sqrt{\inducedmetric}}{\lapse}T^A{}_{ij}T^B{}_{kl}\shift^i\left[\frac{1}{2}M^{j \ l}_{ \ A \ B}\shift^k+\frac{\lapse \phi}{\kappa}\inducedmetric^{jl}\left(c_2 \normalvector_B \tetrad_A{}^k +c_3 \normalvector_A \tetrad_B{}^k \right)\right]+\frac{\lapse \sqrt{\inducedmetric}}{2\kappa} {}^3 \mathbb{T}-\frac{\tetrad V(\phi)}{2\kappa},
	\end{split}
	\end{align}
	with
	\begin{align}
	{}^3 \mathbb{T}=H_{AB}{}^{ijkl}T^A{}_{ij}T^B{}_{kl}=\phi \left(c_1 \eta_{AB}\inducedmetric^{k[i}\inducedmetric^{j]l}-c_2\tetrad_B{}^{[i}\inducedmetric^{j][k}\tetrad_A{}^{l]}-c_3 \tetrad_A{}^{[i}\inducedmetric^{j][k}\tetrad_B{}^{l]} \right)T^A{}_{ij}T^B{}_{kl}.
	\label{eq:TorScalar}
	\end{align}	 
 
 \section{The Hamiltonian for $f(T_\mathrm{NGR})$ gravity}
	\label{sec:Ham}
	
	In order to obtain the Hamiltonian, we need to perform a Legendre transformation to canonical variables and their velocities, with the aim to recast the configuration space in terms of canonical variables and momenta, resulting in the set $(\lapse,\shift^i, \tetrad^A{}_i,\Lambda_A{}^B, \phi,{}^\lapse \pi, {}^\shift \pi_i,\momenta_A{}^i,P^A{}_B, {}^\phi \pi)$, where 
	\begin{eqnarray}
&&	{}^\lapse \pi :=\frac{\partial L}{\partial \dot{\lapse}} = 0,
	\\
&&	{}^\shift \pi_i :=\frac{\partial L}{\partial \dot{\shift}} = 0,
\\
&&	\pi_A{}^i:=\frac{\partial L}{\partial \dot{\tetrad}^A{}_i}:=\frac{\partial L}{\partial T^A{}_{0i}} \nonumber \\ 
	&& \qquad \,  =\frac{\sqrt{\inducedmetric}}{\lapse}M^{i \ j}_{\ A \ B}T^B{}_{0j}-\frac{\sqrt{\inducedmetric}}{\lapse}T^B{}_{kl}\left[M^{i \ l}{}_{\ A \ B}\shift^k+\frac{\lapse\phi}{\kappa}\inducedmetric^{il}\left(c_2 \normalvector_B \tetrad_A{}^k+c_3 \normalvector_A \tetrad_B{}^k \right) \right],
\\
	&& P^A{}_B := \frac{\partial L}{\partial \dot{\Lambda}_A{}^B}=\pi_C{}^i \eta_{AD}\left(\Lambda^{-1} \right)_E{}^B \eta^{C[E}\tetrad^{D]}{}_i,
	\\
	&& {}^\phi \pi := \frac{\partial L}{\partial \dot{\phi}} = 0.
	\end{eqnarray}
	It is worth noticing  the presence of primary constraints, \emph{i.e.}
	\begin{align}
	{}^\lapse C={}^\lapse \pi \approx 0, \quad {}^\shift C_i={}^\shift \pi_i \approx 0, \quad 
	{}^\phi C={}^\phi \pi \approx 0,
	\end{align}
	\begin{align}
	\label{eq:LConstraints}
	{}^\omega C^{AB}= P^{[A}{}_D \eta^{B]C}\Lambda_C{}^D+\pi_C{}^i\eta^{C[B}\tetrad^{A]}{}_i \approx 0,
	\end{align}
	where Eq. \eqref{eq:LConstraints} was found in \cite{Blixt:2018znp,Blixt:2019mkt} by means of the auxiliary fields and later in \cite{Golovnev:2021omn} without the use of auxiliary fields. Among all the $f(T_\mathrm{NGR})$ models, there are several subcases that are fundamentally different, as they realize different symmetries. In the presence of symmetries, there is implicitly a presence of primary constraints, indicated by the fact that the determinant of the Hessian vanishes identically. In NGR, this can be achieved by decomposing velocities and momenta into the irreducible parts under the rotation group. In this way, it is possible to exactly obtain the eigenvalues of the Hessian and, hence, also the primary constraints \cite{Blagojevic:2000qs,Blixt:2020ekl}. The irreducible parts are $\mathcal{V}$ector, $\mathcal{A}$ntisymmetric, $S$ymmetric and trace-free, and $\mathcal{T}$race parts. In short, they constitute the so called $\mathcal{VAST}$ decomposition, which will be adopted from now on in the derivation of the Hamiltonian for $f(T_\mathrm{NGR})$. See also Ref.\cite{Capozziello:2001mq} for a discussion on the decomposition of torsion starting from tetrads and bivectors.
    
    Tetrad fields and momenta, therefore, can be decomposed in terms of such irreducible parts, as:
	\begin{align}
	\dot{\tetrad}^A{}_i={}^{\mathcal{V}}{}\dot{\tetrad}_i\normalvector^A+{}^{\mathcal{A}}{} \dot{\tetrad}_{ji}\inducedmetric^{kj}\tetrad^A{}_k+{}^{\mathcal{S}}{} \dot{\tetrad}_{ji}\inducedmetric^{kj}\tetrad^A{}_k+{}^{\mathcal{T}}{} \dot{\tetrad}\tetrad^A{}_i,
	\label{eq:velocVAST}
	\end{align}
	\begin{align}
	\momenta_A{}^i={}^{\mathcal{V}}{}\momenta^i\normalvector_A+{}^{\mathcal{A}}{} \momenta^{ji}\tetrad^B{}_j\eta_{AB}+{}^{\mathcal{S}}{} \momenta^{ji}\tetrad^B{}_j \eta_{AB}+{}^{\mathcal{T}}{} \momenta\tetrad^B{}_j \eta_{AB}\inducedmetric^{ij}.
	\label{eq:momVAST}
	\end{align}
	
	In order to easily convert our results to the standard variables, we present here the inverse relations
	\begin{align}     
	\begin{split}     
	{}^{\mathcal{S}}{} \dot{\tetrad}_{ji}&=\dot{\tetrad}_{(ji)}- \frac{1}{3}\dot{\tetrad}^A{}_k\tetrad^B{}_l\eta_{AB}\inducedmetric^{kl}\inducedmetric_{ij}=\frac{1}{2} \dot{\tetrad}^A{}_i\tetrad^B{}_j\eta_{AB} +\frac{1}{2}\dot{\tetrad}^A{}_j\tetrad^B{}_i\eta_{AB} -\frac{1}{3}\dot{\tetrad}^A{}_k\tetrad^B{}_l\eta_{AB}\inducedmetric^{kl}\inducedmetric_{ij}, \\
	{}^{\mathcal{T}}{} \dot{\tetrad}&=\frac{1}{3} \dot{\tetrad}^A{}_i \tetrad^B{}_j\eta_{AB}\inducedmetric^{ij},\\
	{}^{\mathcal{V}}{} \dot{\tetrad}_i & =-\normalvector_A \dot{\tetrad}^A{}_i,\\
	{}^{\mathcal{A}}{} \dot{\tetrad}_{ji} & = \dot{\tetrad}_{[ji]}=\frac{1}{2} \dot{\tetrad}^A{}_i \tetrad^B{}_j\eta_{AB} -\frac{1}{2}\dot{\tetrad}^A{}_j \tetrad^B{}_i\eta_{AB},
	\label{eq.27}
	\end{split} \end{align}
	and
	\begin{equation}
	\begin{split}
	{}^{\mathcal{S}}\momenta^{ji}&=\momenta^{(ji)}-\frac{1}{3}\momenta_A{}^k\tetrad^A{}_k\inducedmetric^{ij}=\frac{1}{2}\momenta_A{}^i\tetrad^A{}_k\inducedmetric^{jk}+\frac{1}{2}\momenta_A{}^j\tetrad^A{}_k\inducedmetric^{ik}-\frac{1}{3}\momenta_A{}^k\tetrad^A{}_k\inducedmetric^{ij}, \\
	{}^{\mathcal{T}} \momenta&=\frac{1}{3} \momenta_A{}^i \tetrad^A{}_i,\\
	{}^{\mathcal{V}} \momenta^i &=-\normalvector^A \momenta_A{}^i,\\
	{}^{\mathcal{A}} \momenta^{ji} &=\momenta^{[ji]}=\frac{1}{2}\momenta_A{}^i \tetrad^A{}_k\inducedmetric^{jk}-\frac{1}{2}\momenta_A{}^j \tetrad^A{}_k\inducedmetric^{ik}.
	\end{split}
	\label{symtrcMomenta}
	\end{equation}
	Also the Hessian, which is contracted with the velocities, can be recast in the $\mathcal{VAST}$ decomposition as:
	\begin{align}
	\begin{split}
	M^{i \ j}_{\ A \ B}&={}^\mathcal{V}M^{ij}\normalvector_A \normalvector_B+{}^\mathcal{A}M^{[ik][jl]}\tetrad^C{}_k \eta_{AC} \tetrad^D{}_l \eta_{BD}+{}^\mathcal{S}M^{(ik)(jl)}\tetrad^C{}_k \eta_{AC} \tetrad^D{}_l\eta_{BD}+{}^\mathcal{T}M \tetrad_A{}^i \tetrad_B{}^j.
	\end{split}
	\end{align}
	For our general discussion of $f(T_\mathrm{NGR})$ gravity, we collect the $\mathcal{VAST}$ labels as $\mathcal{I} \in \{\mathcal{V},\mathcal{A},\mathcal{S},\mathcal{T}\}$. With the aim to get a vanishing determinant for the Hessian $M^{i \ j}_{\ A \ B }$, when $\mathcal{A}_\mathcal{I}=0$, we consider the following relations:
	\begin{align}
	\mathcal{A}_\mathcal{V}&=2c_1+c_2+c_3, \\
	\mathcal{A}_\mathcal{A}&=2c_1-c_2, \\
	\mathcal{A}_\mathcal{S}&=2c_1+c_2, \\
	\mathcal{A}_\mathcal{T}&=2c_1+c_2+3c_3.
	\end{align}
	The list of possible constraints in $f(T_\mathrm{NGR})$, coinciding with $\mathcal{A}_\mathcal{I}=0 \implies {}^\mathcal{I}C \approx 0$,  is 
	\begin{align} 
	{}^{\mathcal{V}}C^i =  \frac{{}^{\mathcal{V}} \pi^i\kappa}{\phi \sqrt{\inducedmetric}}+c_3 T^B{}_{jk}\inducedmetric^{ik}\inducedmetric^{jl}\tetrad^A{}_l\eta_{AB}\approx 0, \label{Cvek}
	\end{align}
	\begin{align}
	{}^{\mathcal{A}}C^{ij} =  \frac{{}^{\mathcal{A}} \pi^{ij}\kappa}{\phi \sqrt{\inducedmetric}}+c_2 \inducedmetric^{ik}\inducedmetric^{jl}T^B{}_{kl}\normalvector_B\approx 0,\label{Casy}
	\end{align}
 
	\begin{align}
	{}^{\mathcal{S}}C^{ij} =  \frac{{}^{\mathcal{S}} \pi^{ij}\kappa}{\phi \sqrt{\inducedmetric}}\approx 0,\label{Csy}
	\end{align}
	\begin{align}
	{}^{\mathcal{T}}C =  \frac{{}^{\mathcal{T}} \pi\kappa}{\phi \sqrt{\inducedmetric}}\approx 0.\label{Ctr}
	\end{align}
	To complete the Legendre transformation from the Lagrangian to the Hamiltonian, the velocities need to be rewritten in terms of canonical Hamiltonian variables. To this purpose, it is necessary to find the Moore-Penrose pseudoinverse of the Hessian, which can also be written in the $\mathcal{VAST}$ decomposition as: 
	\begin{align}
	\begin{split}
	\left(M^{-1} \right)^{\ A \ C}_{ i \ k}&=\frac{\kappa}{\phi}\mathcal{B}_\mathcal{V}\normalvector^A \normalvector^C  \inducedmetric_{ik}-\frac{\kappa}{\phi}\mathcal{B}_{\mathcal{A}}\inducedmetric^{r[s}\inducedmetric^{m]n}\inducedmetric_{kr}\inducedmetric_{si} \tetrad^A{}_m \tetrad^C{}_n\\
	&-\frac{\kappa}{\phi} \mathcal{B}_\mathcal{S}\left(\inducedmetric^{r(s}\inducedmetric^{m)n}-\frac{1}{3}\inducedmetric^{sm}\inducedmetric^{nr} \right)\inducedmetric_{kr}\inducedmetric_{si}\tetrad^A{}_m \tetrad^C{}_n-\frac{\kappa}{3\phi} \mathcal{B}_\mathcal{T}\tetrad^A{}_i \tetrad^C{}_k,
	\end{split}
	\end{align}
	with 
	\begin{align}
	\mathcal{B}_\mathcal{I}=\begin{cases}
	\frac{1}{\mathcal{A}_\mathcal{I}}, \indent \mathrm{if} \ \mathcal{A}_\mathcal{I}\neq 0,\\
	0, \ \ \  \indent \mathrm{if} \ \mathcal{A}_\mathcal{I}=0
	\end{cases}.
	\end{align}
	The velocities can, thus, be expressed in canonical Hamiltonian variables:
	\begin{align}
	\begin{split}
	T^C{}_{0k}=\left(M^{-1} \right)^{\ A \ C}_{i \ k}\frac{\alpha}{\sqrt{\inducedmetric}}\pi_A{}^i+T^C{}_{mk}\shift^m-\frac{\lapse}{\kappa}\left(M^{-1} \right)^{\ A \ C}_{i \ k}T^B{}_{ml}\inducedmetric^{il}\left(\frac{1}{2}\normalvector_B \tetrad_A{}^m -\normalvector_A \tetrad_B{}^m \right),
	\end{split}
	\end{align}
	so that
	\begin{align}
	\begin{split}
	\dot{\tetrad}^C{}_{k}-\left(\Lambda^{-1} \right)^A{}_B \dot{\Lambda}_A{}^C\tetrad^B{}_k&=\partial_k\tetrad^C{}_0+\spinconnection^C{}_{Dk}\tetrad^D{}_0+\left(M^{-1} \right)^{\ A \ C}_{i \ k}\frac{\alpha}{\sqrt{\inducedmetric}}\pi_A{}^i+T^C{}_{mk}\shift^m\\
	&-\frac{\lapse}{\kappa}\left(M^{-1} \right)^{\ A \ C}_{i \ k}T^B{}_{ml}\inducedmetric^{il}\left(\frac{1}{2}\normalvector_B \tetrad_A{}^m -\normalvector_A \tetrad_B{}^m \right).
	\end{split}
	\end{align}
	The covariant formulation and the primary constraints associated with this formalism require the tetrad and Lorentz matrix velocities to be inverted together. The Hamiltonian density, in this way, is given by
	\begin{align}
	\mathcal{H}_c=  \momenta_A{}^i\left(\dot{\tetrad}^A{}_{k}-\left(\Lambda^{-1} \right)^C{}_B \dot{\Lambda}_C{}^A\tetrad^B{}_i\right)-\mathcal{L},
	\end{align}
	which explicitly reads
	\begin{align}
 \label{eq:HfTNGR}
	\begin{split}
	\mathcal{H}_{f(T_\mathrm{NGR})}&=\lapse  \left[ \frac{\sqrt{\inducedmetric}\phi}{2\kappa} \mathcal{B}_\mathcal{V}{}^{\mathcal{V}}C^i {}^{\mathcal{V}} C_i-\frac{\sqrt{\inducedmetric}\phi}{2\kappa} \mathcal{B}_\mathcal{A}{}^{\mathcal{A}}C^{ij} {}^{\mathcal{A}} C_{ij}-\frac{\sqrt{\inducedmetric}\phi}{2\kappa} \mathcal{B}_\mathcal{S}{}^{\mathcal{S}}C^{ij} {}^{\mathcal{S}} C_{ij}-\frac{3\sqrt{\inducedmetric}\phi}{2\kappa} \mathcal{B}_\mathcal{T}{}^{\mathcal{T}}C {}^{\mathcal{T}} C\right. \\& \left.-\frac{\sqrt{\inducedmetric}}{2\kappa}{}^3 \mathbb{T}+\frac{\sqrt{\inducedmetric}V(\phi)}{2\kappa}-\normalvector^A\partial_i \momenta_A{}^i+\momenta_A{}^i\spinconnection^A{}_{Bi}\normalvector^B \right]  \\
	&+\shift^j\left[ -\tetrad^A{}_j \partial_i \momenta_A{}^i+\momenta_A{}^i\spinconnection^A{}_{Ci} \tetrad^C{}_j-\momenta_A{}^i T^A{}_{ij} \right] -{}^\alpha \lambda {}^\alpha \pi-{}^\beta \lambda_i {}^\beta \pi^i-{}^\phi \lambda {}^\phi \pi \\
	&-\lambda_{AB}\left( P^{[A}{}_D\eta^ {B]C}\Lorentz_C{}^D+\momenta_C{}^i\eta^{C[B}\tetrad^{A]}{}_i \right)-{}^\mathcal{S}\lambda_{ij}\frac{{}^\mathcal{S} \pi^{ij}\kappa}{\phi \sqrt{\gamma}}-{}^\mathcal{T}\lambda \frac{{}^\mathcal{T}\pi \kappa}{\phi \sqrt{\inducedmetric}} \\
	&-{}^{\mathcal{V}}\lambda_i\left(\frac{{}^{\mathcal{V}} \pi^i\kappa}{\sqrt{\inducedmetric}}+c_3 T^B{}_{jk}\inducedmetric^{ik}\inducedmetric^{jl}\tetrad^A{}_l\eta_{AB}\right)-{}^{\mathcal{A}}\lambda_{ij}\left(\frac{{}^{\mathcal{A}} \pi^{ij}\kappa}{\sqrt{\inducedmetric}}+c_2  \inducedmetric^{ik}\inducedmetric^{jl}T^B{}_{kl}\normalvector_B \right)\\
	&+\partial_i \left(\momenta_A{}^i\tetrad^A{}_0 \right),
	\end{split}
	\end{align}
	where ${}^\mathcal{V}\lambda_i = 0$, unless ${}^\mathcal{V}\mathcal{A}=0$ and similarly for ${}^\mathcal{A}\lambda_{ij} $, ${}^\mathcal{S}\lambda_{ij} $ and ${}^\mathcal{T}\lambda $, which vanish as well unless the corresponding ${}^\mathcal{A}\mathcal{A}$, ${}^\mathcal{S}\mathcal{A}$ or ${}^\mathcal{T}\mathcal{A}$ are zero.
	The boundary term $\partial_i \left(\momenta_A{}^i\tetrad^A{}_0 \right)$ contains non-linearity in lapse and shift, which spoils the Hamiltonian and momenta constraints. Thus, this boundary term will be dropped for the rest of the paper as was done in \cite{Arnowitt:1962hi,Pati:2022nwi}.
	
 To simplify the derivation of  Hamilton equations, it is worth noticing  that the symmetric trace-free and trace part can be combined with the symmetric part in the following way
\begin{align}
\label{eq:momsquared}
    \begin{split}
        -\frac{\lapse \sqrt{\inducedmetric}\phi}{2\kappa}\left(\mathcal{B}_\mathcal{S} {}^{\mathcal{S}}C^{ij}{}^{\mathcal{S}}C_{ij}+3\mathcal{B}_\mathcal{T} {}^{\mathcal{T}}C{}^{\mathcal{T}}C \right)&=-\frac{\kappa \mathcal{B}_\mathcal{S}}{4\sqrt{\inducedmetric}\phi}\left(\momenta_A{}^i\momenta_B{}^l\tetrad^A{}_k\tetrad^B{}_j\inducedmetric^{jk}\inducedmetric_{il}+\momenta_A{}^i\momenta_B{}^k\tetrad^A{}_k\tetrad^B{}_i \right) \\
        &+\frac{\kappa(\mathcal{B}_\mathcal{S}-\mathcal{B}_\mathcal{T})}{6\sqrt{\inducedmetric}\phi}\momenta_A{}^i \momenta_B{}^j\tetrad^A{}_i\tetrad^B{}_j.
    \end{split}
\end{align}
In sensible teleparallel theories, $\mathcal{B}_\mathcal{S} =-1$ and $\mathcal{B}_\mathcal{T}=\frac{1}{2}$, which is the case of TEGR, $f(T)$ and NGR, giving the expected propagation of a massless spin-2 field. From this expression, it is straightforward to find consistency with the Hamiltonians typically presented in teleparallel theories \cite{Cheng:1988zg,Li:2011rn,Okolow:2011nq,Maluf:2013gaa,Blagojevic:2020dyq,Pati:2022nwi,Blixt:2020ekl}. In addition, also the derivation of the Hamilton equations is less cumbersome starting from this expression. Although so far for simplicity we have dealt with the Hamiltonian density $\mathcal{H}$, to get the expression for the Hamilton equations, hereafter we will consider the Hamiltonian $H=\int d^3x \mathcal{H}$.

	\section{The Hamilton equations for $f(T)$ teleparallel gravity}
 \label{sec:HeqfT}
	
	From the perspective of $f(T_\mathrm{NGR})$ gravity, $f(T)$ gravity can be obtained when fixing the coefficients appearing in the torsion scalar \eqref{TNGR} to those of TEGR. This implies that $\mathcal{B}_\mathcal{V}=\mathcal{B}_\mathcal{A}={}^\mathcal{S}\lambda_{ij}={}^{\mathcal{T}}\lambda=0$, so that ${}^\mathcal{V}\lambda_i \neq 0$ and ${}^\mathcal{A}\lambda_{ij} \neq 0$. Furthermore one must also impose $\mathcal{B}_\mathcal{S}=-1$,  $\mathcal{B}_{\mathcal{T}}=\dfrac{1}{2}$, $c_1=-\dfrac{1}{4}$, $c_2=-\dfrac{1}{2}$, and $c_3=1$. The theory contains primary constraints that are not of first class \cite{Blagojevic:2020dyq}. Since Hamilton's equations are valid only on the constraint surface \cite{Bojowald_2010}, it is important to derive all constraints in the evaluation, which is equivalent to demanding that the constraints are preserved in time.
 	
	\subsection{Without gauge fixing}
  \label{sec:HeqfTw}
	Even though it is known that the Weitzenböck gauge can always be chosen consistently \cite{Golovnev:2021omn,Blixt:2022rpl,Golovnev:2023yla}, there are still reasons to consider the covariant formulation. Firstly, not all observers admit the foliation (see Ref. \cite{Borowiec:2013kgx,Blixt:2024aej}) assumed in Sec. \ref{sec:3+1}. In this context, the spin-connection might play an important role to guarantee foliation. Another motivation is related to avoiding a divergent boundary term \cite{Gomes:2022vrc}. Finally, not all gauges are suitable for numerical relativity \cite{Baumgarte:2002jm}. Therefore, it is convenient to present the  Hamilton equations in the covariant formulation, since the Weitzenböck gauge may result in an impractical choice for these applications. As mentioned above, the Hamiltonian for $f(T)$ gravity can be obtained from Eq. \eqref{eq:HfTNGR}, by properly choosing the coefficients $c_i$. It is given by the expression
 \begin{align}
	\begin{split}
	\mathcal{H}_{f(T)}&=\lapse  \left[ \frac{\kappa}{4\sqrt{\inducedmetric}\phi}\left(2\momenta_A{}^i\momenta_B{}^l\tetrad^A{}_j\tetrad^B{}_{(k}\inducedmetric_{i)l}\inducedmetric^{jk}-\momenta_A{}^i\momenta_B{}^j\tetrad^A{}_i\tetrad^B{}_j\right) -\frac{\sqrt{\inducedmetric}}{2\kappa}{}^3 \mathbb{T}+\frac{\sqrt{\inducedmetric}V(\phi)}{2\kappa}-\normalvector^A\partial_i \momenta_A{}^i\right. \\
 &\left.+\momenta_A{}^i\spinconnection^A{}_{Bi}\normalvector^B \right]  +\shift^j\left[ -\tetrad^A{}_j \partial_i \momenta_A{}^i+\momenta_A{}^i\spinconnection^A{}_{Ci} \tetrad^C{}_j-\momenta_A{}^i T^A{}_{ij} \right] -{}^\alpha \lambda {}^\alpha \pi-{}^\beta \lambda_i {}^\beta \pi^i-{}^\phi \lambda {}^\phi \pi \\
	&-{}^{\mathcal{V}}\lambda_i\left(\frac{{}^{\mathcal{V}} \pi^i\kappa}{\sqrt{\inducedmetric}}+T^B{}_{jk}\inducedmetric^{ik}\inducedmetric^{jl}\tetrad^A{}_l\eta_{AB}\right)-{}^{\mathcal{A}}\lambda_{ij}\left(\frac{{}^{\mathcal{A}} \pi^{ij}\kappa}{\sqrt{\inducedmetric}}-\frac{1}{2}  \inducedmetric^{ik}\inducedmetric^{jl}T^B{}_{kl}\normalvector_B \right)\\
	&-\lambda_{AB}\left( P^{[A}{}_D\eta^ {B]C}\Lorentz_C{}^D+\momenta_C{}^i\eta^{C[B}\tetrad^{A]}{}_i \right) .
	\end{split}
	\end{align}
 Below, we present Hamilton's equations for covariant $f(T)$ gravity. First, let us start by considering the following Hamiltonian constraints:
	
	\begin{align}
		\begin{split}
		-{}^{\lapse}\dot{\momenta}&=\frac{\delta H}{\delta \lapse}=\frac{\kappa}{4\sqrt{\inducedmetric}\phi}\left(2\momenta_A{}^i\momenta_B{}^l\tetrad^A{}_j\tetrad^B{}_{(k}\inducedmetric_{i)l}\inducedmetric^{jk}-\momenta_A{}^i\momenta_B{}^j\tetrad^A{}_i\tetrad^B{}_j\right) -\frac{\sqrt{\inducedmetric}}{2\kappa}{}^3 \mathbb{T}+\frac{\sqrt{\inducedmetric}V(\phi)}{2\kappa}-\normalvector^A\partial_i \momenta_A{}^i\\
  &+\momenta_A{}^i\spinconnection^A{}_{Bi}\normalvector^B. 
		\end{split}
	\end{align}
	The momenta constraint does not depend on the specific teleparallel theory, as seen by comparison with Eq. \eqref{eq:momconstfTNGR} below. Nevertheless, it still depends on the gauge choice, as it can be inferred by the presence of the spin-connection. It reads:
		\begin{align}
	\begin{split}
	-{}^{\shift}\dot{\momenta}_i&=\frac{\delta H}{\delta \shift^i}=-\tetrad^A{}_i \partial_j \momenta_A{}^j+\momenta_A{}^j\spinconnection^A{}_{Cj} \tetrad^C{}_i-\momenta_A{}^j T^A{}_{ji}.
	\end{split}
	\end{align}
	Similarly to the case of TEGR \cite{Pati:2022nwi}, the time evolutions of the conjugate momenta $\dot{\momenta}_A{}^i$ and $P^A{}_B$ are very lengthy and are given by the expressions below:
 				\begin{align}
     \label{momentalarge}
	\begin{split}
	-\dot{\momenta}_A{}^i&=\frac{\delta H}{\delta \tetrad^A{}_i}=\lapse \left(-\frac{2\sqrt{\inducedmetric}}{\kappa}H_{CB}{}^{mn[ki]}T^C{}_{mn}\omega^B{}_{Ak}+\frac{\sqrt{\inducedmetric}}{2\kappa}\tetrad_A{}^i\left(V(\phi)-{}^3 \mathbb{T} \right) \right. \\
	& \left. -\frac{\sqrt{\inducedmetric}}{\kappa}T^C{}_{mn}T^B{}_{kl}\left(\tetrad_A{}^m H_{CB}{}^{inlk}+\tetrad_A{}^k H_{CB}{}^{mnli}\right. \right. \\
	&\left. \left. -\frac{1}{2}\normalvector_A\normalvector_C\tetrad_B{}^{[m}\inducedmetric^{n][k}\inducedmetric^{l]i}+\normalvector_A \normalvector_B\tetrad_C{}^{[m}\inducedmetric^{n][k}\inducedmetric^{l]i} \right)+\inducedmetric^{im}\normalvector_A\tetrad^C{}_m\left(\partial_j \momenta_C{}^j-\momenta_B{}^j\omega^B{}_{Cj}\right) \right) \\
	& +\shift^j\left( \momenta_B{}^{i}\omega^B{}_{Aj}-\delta_j^i\partial_k\momenta_A{}^k \right)+\lambda_{[BA]}\eta^{BC}\momenta_C{}^i\\
	&+\partial_j \left(\frac{2\lapse\sqrt{\inducedmetric}}{\kappa}H_{BA}{}^{kl[ij]}T^B{}_{kl}+2\shift^{[i}\momenta_A{}^{j]}+2{}^{\mathcal{V}}\lambda^{[i} \tetrad_A{}^{j]}+{}^{\mathcal{A}}\lambda^{[ij]} \normalvector_A \right)\\
	&+\frac{\lapse\kappa}{2\sqrt{\inducedmetric}\phi} \left(\momenta_C{}^{(i}\momenta_B{}^{l)}\tetrad^C{}_j\tetrad^B{}_k\eta_{AD}\inducedmetric^{jk}\tetrad^D{}_l-\momenta_C{}^m\momenta_B{}^l\tetrad^C{}_j\tetrad^B{}_k\tetrad_A{}^{(j}\inducedmetric^{k)i}\inducedmetric_{ml}+2\momenta_A{}^j\momenta_B{}^{(k}\inducedmetric^{i)l}\inducedmetric_{jk}\tetrad^B{}_l  \right. \\
 &\left. -\momenta_A{}^i\momenta_B{}^j\tetrad^B{}_j-\tetrad_A{}^i\momenta_C{}^m\momenta_B{}^{(l}\inducedmetric^{k)j}\tetrad^C{}_k\tetrad^B{}_j\inducedmetric_{ml}+\frac{1}{2}\tetrad_A{}^i\momenta_C{}^k\momenta_B{}^j\tetrad^C{}_k\tetrad^B{}_j\right)   \\
	&+{}^{\mathcal{V}}\lambda_j \left(-2\inducedmetric^{j[i}\tetrad_B{}^{k]}\omega^B{}_{Ak}-\frac{\kappa}{\sqrt{\inducedmetric}\phi}\inducedmetric^{im}\normalvector_A\tetrad^C{}_m\momenta_C{}^j-  \inducedmetric^{ik}\inducedmetric^{jl}\eta_{AB}T^B{}_{kl} \right. \\
	& \left. -\frac{\kappa}{\sqrt{\inducedmetric}\phi}\normalvector^B\momenta_B{}^j\tetrad_A{}^i+2 T^B{}_{kl}\tetrad_B{}^{[i}\inducedmetric^{j]k} \tetrad_A{}^l+2  \inducedmetric^{il}\eta_{BC}T^B{}_{kl}\tetrad^C{}_m\tetrad_A{}^{[m}\inducedmetric^{j]k}
	 \right)\\
	 &+{}^{\mathcal{A}}\lambda_{[jk]} \left(\inducedmetric^{kl}\inducedmetric^{ij}\normalvector_B\omega^B{}_{Al}+\frac{1}{2}\normalvector_A\tetrad_B{}^i\inducedmetric^{jn}\inducedmetric^{km}T^B{}_{nm}+\frac{\kappa}{\sqrt{\inducedmetric}\phi}\inducedmetric^{jl}\momenta_B{}^k\tetrad^B{}_l\tetrad_A{}^i \right. \\
	 &\left. - \inducedmetric^{i[k}\inducedmetric^{j]m}\normalvector_B T^B{}_{ml}\tetrad_A{}^l- \inducedmetric^{im}\normalvector_B T^B{}_{nm}\tetrad_A{}^{[k}\inducedmetric{}^{j]n}+\frac{2\kappa}{\sqrt{\inducedmetric}\phi} \momenta_B{}^k \tetrad^B{}_l\tetrad_A{}^{[l}\inducedmetric^{j]i}+\frac{\kappa}{\sqrt{\inducedmetric}\phi}\inducedmetric^{ik}\momenta_A{}^j \right),
	\end{split}
	\end{align}
and
 \begin{align}
 \label{Plarge}
\begin{split}
-\dot{P}^A{}_B &= \dfrac{\delta H}{\delta \Lambda_{A}{}^{B} }  =
\lambda_{[DC]}\eta^{AD}P^C{}_B-\lapse \left( \Lambda^{-1}\right)^A{}_D\omega^C{}_{Bi}\normalvector^D\momenta_C{}^i-\shift^{i} \momenta_C{}^{j} \left(\Lambda^{-1} \right)^A{}_D \omega^C{}_{Bi}  \tetrad^D{}_j \\
&-(H_{CE}{}^{kijl}-H_{CE}{}^{iljk})\frac{\lapse \sqrt{\inducedmetric} \left(\Lambda^{-1} \right)^A{}_D \omega^C{}_{Bi} T^E{}_{jl}\tetrad^D{}_k}{\kappa} +{}^\mathcal{A}\lambda^{[ij]} \left(\Lambda^{-1} \right)^A{}_D \omega^C{}_{Bi} \normalvector_C  \tetrad^D{}_j \\
&-2{}^{\mathcal{V}}\lambda_k  \left(\Lambda^{-1} \right)^A{}_D \omega^C{}_{Bi} \tetrad_C{}^{[i}\inducedmetric^{k]j} \tetrad^D{}_j   +\partial_i \left[\lapse \left(\Lambda^{-1} \right)^A{}_D \normalvector^D \pi_B{}^i-\shift^i \left(\Lambda^{-1} \right)^A{}_D \momenta_B{}^k \right. \\
& \left. +\frac{\lapse\sqrt{\inducedmetric} }{\kappa}\left(H_{BC}{}^{kijl}-H_{BC}{}^{iljk}\right)\left(\Lambda^{-1} \right)^A{}_D T^C{}_{jl} \tetrad^D{}_k  -{}^\mathcal{A}\lambda^{[ij]} \left(\Lambda^{-1} \right)^A{}_D  \normalvector_B \tetrad^D{}_j\right. \\
& \left. +2{}^\mathcal{V}\lambda_l  \left(\Lambda^{-1} \right)^A{}_D \inducedmetric^{i[j}\inducedmetric^{l]k}\eta_{BC} \tetrad^C{}_j\tetrad^D{}_k \right].
\end{split}
\end{align}
It is worth stressing that the dependence of above equation on the scalar field $\phi$ is implicitly contained in the super-metric $H_{AB}{}^{ijkl}(\phi)$.


 
 The completely novel result is the time evolution of the conjugate momenta of the scalar field, reading as:
	
		\begin{align}
	\begin{split}
	-{}^{\phi}\dot{\momenta}&=\frac{\delta H}{\delta \phi}=\lapse\left[  -\frac{\kappa }{2\sqrt{\inducedmetric} \phi^2}\left(\inducedmetric^{kl}\inducedmetric_{i(j} \tetrad^A{}_{k)} \tetrad^B{}_l\momenta_A{}^i \momenta_B{}^j  -\frac{\momenta_A{}^i\momenta_B{}^j\tetrad^A{}_i \tetrad^B{}_j }{2}\right) +\frac{ \sqrt{\inducedmetric}}{2\kappa}\frac{\delta V(\phi)}{\delta \phi}-\frac{\sqrt{\inducedmetric}}{2\kappa \phi}{}^3 \mathbb{T} \right]\\
	&-{}^\mathcal{V}\lambda_i \frac{\kappa \normalvector^A \momenta_A{}^i }{\sqrt{\inducedmetric} \phi^2}+{}^\mathcal{A}\lambda_{[ik]}\frac{\kappa \inducedmetric^{ij}\momenta_A{}^k \tetrad^A{}_j}{\sqrt{\inducedmetric} \phi^2}
	\end{split}
	\end{align}
	This result is similar to the case of $f(R)$ gravity, (see Eq. (3.10) in \cite{Deruelle:2009zk}). Moving on to the evolution of the canonical variables, it is simple to find that the time evolution of lapse function and shift vector, that is $\dot{\lapse}$ and $\dot{\shift}^i$, are proportional to the Lagrange multipliers:
	\begin{align}
		\begin{split}
		\dot{\lapse}=\frac{\delta H}{\delta {}^{\lapse}\momenta}=-{}^{\lapse}\lambda,
		\end{split}
	\end{align}
	
		\begin{align}
	\dot{\beta}^i=\frac{\delta H}{\delta {}^\beta\pi_i}=-{}^\shift \lambda^i.
	\end{align}
 The time evolution of the tetrad, instead, reads as:
	\begin{align}
	\begin{split}
	\dot{\theta}^A{}_i=\frac{\delta H}{\delta \pi_A{}^i}&= \lapse\left[
	 \frac{ \kappa}{\sqrt{\inducedmetric} \phi }\left(\inducedmetric^{kl}\inducedmetric_{i(j}\tetrad^A{}_{k)}\tetrad^B{}_l \momenta_B{}^j -\frac{\momenta_B{}^j\tetrad^A{}_i \tetrad^B{}_j }{2}\right)+\partial_i\normalvector^A+\normalvector^B\omega^A{}_{Bi} \right]\\
  &+\shift^j\left[\tetrad^B{}_j \omega^A{}_{Bi}-T^A{}_{ij}\right]+ \lambda_{[CB]}\tetrad^B{}_i\eta^{AC}+{}^\mathcal{V}\lambda_i \frac{\kappa \normalvector^A}{\sqrt{\inducedmetric} \phi}+{}^\mathcal{A}\lambda_{[ik]}\frac{\kappa\inducedmetric^{jk}\tetrad^A{}_j}{\sqrt{\inducedmetric} \phi}+\partial_i\left(\shift^j \tetrad^A{}_j \right).
	\end{split}
	\end{align}
 Notice that the above equation has an important difference from TEGR, since it involves the scalar field which spoils the Lorentz symmetry imposed by the primary constraints associated with the Lagrange multipliers ${}^\mathcal{V}\lambda_i$ and ${}^\mathcal{A}\lambda_{ij}$ \cite{Blagojevic:2020dyq}. The time evolution of the Lorentz matrices is only related to the primary constraints associated with the Lorentz covariance, which implies transforming the tetrad and spin-connection together (defined as Lorentz covariance of type II in \cite{Blixt:2022rpl})
	\begin{align}
 \label{lambdadot}
	\begin{split}
	\dot{\Lambda}_A{}^B&=\frac{\delta H}{\delta P^A{}_B}= \lambda_{[DA]}\Lambda_C{}^B \eta^{CD}.
	\end{split}
	\end{align}
	Thus, Eq. \eqref{lambdadot} reads the same in all teleparallel theories, as can be noticed by comparing it to Eq. \eqref{eq:LorentzEv}. Lastly, as is well-known, the time evolution of the scalar field is proportional to a Lagrange multiplier opposed to the case of $f(R)$ (which can be understood through Eq. (3.9) in \cite{Deruelle:2009zk}):
	\begin{align}
	\dot{\phi}=\frac{\delta H}{\delta {}^\phi \pi}=-{}^\phi \lambda. 
	\end{align}
	In the limit $\phi=1$ and $V(\phi)=0$, TEGR is straightforwardly recovered. As TEGR results most easy to compare with $f(T)$ gravity, here we check the consistency with Ref. \cite{Pati:2022nwi}. From the comparison, it is easy to see that, when $\phi =1$, the quantities ${}^\lapse \dot{\momenta}, {}^\shift \dot{\momenta}_i, \dot{\lapse},\dot{\shift}^i, \dot{\tetrad}^A{}_i, \dot{\Lambda}_A{}^B$ coincide with those listed in \cite{Pati:2022nwi}. Regarding $\dot{P}_A{}^B$, we find out a typo in \cite{Pati:2022nwi} with a couple of terms having the free index $C$, which should be a free index A instead. Also, in both cases of $\dot{P}^A{}_B$ and $\dot{\momenta}_A{}^i$, we find that the super-metric $H$ cannot be reduced as much as was done in \cite{Pati:2022nwi}. However, on the contrary, we find that the terms including quadratic torsion can be further simplified using the symmetries of the torsion contained in $\dot{\momenta}_A{}^i$ (see Eq. \eqref{eq:Hsimple}). We found a few more sign mistakes for $\dot{\momenta}_A{}^i$ in \cite{Pati:2022nwi} and for this reason we present $\dot{\momenta}_A{}^i$ explicitly for TEGR in Eq. \eqref{eq:TEGRmom}. The derived Hamilton equations, together with the primary constraints, can thus be used to show explicitly if the degrees of freedom for $f(T)$ gravity is indeed the same as in the Weitzenböck gauge. According to the step outlined in \cite{Golovnev:2021omn,Blixt:2018znp,Blixt:2019mkt}, the latter statement can be explicitly proven by evaluating the following relation:

\begin{align}
\label{relatCAB}
    \begin{split}
        {}^\omega \dot{C}^{AB}&=\dot{P}^{[A}{}_D \eta^{B]C}\Lambda_C{}^D+P^{[A}{}_D \eta^{B]C}\dot{\Lambda}_C{}^D+\dot{\pi}_C{}^i\eta^{C[B}\tetrad^{A]}{}_i+\pi_C{}^i\eta^{C[B}\dot{\tetrad}^{A]}{}_i \approx 0.
    \end{split}
\end{align}
Due to the large expressions of $\dot{\momenta}_A{}^i$ and $\dot{P}^A{}_B$, written respectively in Eqs. \eqref{momentalarge} and \eqref{Plarge}, the explicit calculation of Eq. \eqref{relatCAB} would be quite lengthy. This can constitute a strong proof for the viability of the Weitzenböck gauge, where the same expression turns out to be simpler to handle, as pointed out in the next subsection. Nevertheless, as argued before, there are still advantages to deal with the covariant formulation, since several aspects have not been investigated in detail yet, such as the conditions for foliation \cite{Blixt:2024aej} and strong hyperbolicity in $f(T)$ gravity. Moreover, the covariant formulation could be also important in the definition of energy and mass, as indicated in Ref. \cite{Gomes:2022vrc,Gomes:2023hyk}.

 \subsection{The Weitzenböck gauge}
  \label{sec:HeqfTw0}
The Hamiltonian for $f(T)$ gravity simplifies to the following expression in the Weitzenböck gauge 
 \begin{align}
	\begin{split}
	\mathcal{H}_{f(T)}&=\lapse  \left[ \frac{\kappa}{4\sqrt{\inducedmetric}\phi}\left(2\momenta_A{}^i\momenta_B{}^l\tetrad^A{}_j\tetrad^B{}_{(k}\inducedmetric_{i)l}\inducedmetric^{jk}-\momenta_A{}^i\momenta_B{}^j\tetrad^A{}_i\tetrad^B{}_j\right) -\frac{\sqrt{\inducedmetric}}{2\kappa}{}^3 \mathbb{T}+\frac{\sqrt{\inducedmetric}V(\phi)}{2\kappa}-\normalvector^A\partial_i \momenta_A{}^i \right]\\
 &+\shift^j\left[ -\tetrad^A{}_j \partial_i \momenta_A{}^i-\momenta_A{}^i T^A{}_{ij} \right] -{}^\alpha \lambda {}^\alpha \pi-{}^\beta \lambda_i {}^\beta \pi^i-{}^\phi \lambda {}^\phi \pi \\
	&-{}^{\mathcal{V}}\lambda_i\left(\frac{{}^{\mathcal{V}} \pi^i\kappa}{\sqrt{\inducedmetric}}+T^B{}_{jk}\inducedmetric^{ik}\inducedmetric^{jl}\tetrad^A{}_l\eta_{AB}\right)-{}^{\mathcal{A}}\lambda_{ij}\left(\frac{{}^{\mathcal{A}} \pi^{ij}\kappa}{\sqrt{\inducedmetric}}-\frac{1}{2}  \inducedmetric^{ik}\inducedmetric^{jl}T^B{}_{kl}\normalvector_B \right) ,
	\end{split}
	\end{align}
 with the Hamiltonian constraint
		\begin{align}
	\begin{split}
	-{}^{\lapse}\dot{\momenta}&=\frac{\delta H}{\delta \lapse}=\frac{\kappa}{4\sqrt{\inducedmetric}\phi}\left(2\momenta_A{}^i\momenta_B{}^l\tetrad^A{}_j\tetrad^B{}_{(k}\inducedmetric_{i)l}\inducedmetric^{jk}-\momenta_A{}^i\momenta_B{}^j\tetrad^A{}_i\tetrad^B{}_j\right)  -\frac{\sqrt{\inducedmetric}}{2\kappa}{}^3 \mathbb{T}+\frac{\sqrt{\inducedmetric}V(\phi)}{2\kappa}-\normalvector^A\partial_i \momenta_A{}^i,
	\end{split}
	\end{align}
differing from TEGR by the presence of $\phi$, which appears both explicitly and in the definition of ${}^3 \mathbb{T}(\phi)$. The momenta constraint
	\begin{align}
	\begin{split}
	-{}^{\shift}\dot{\momenta}_i&=\frac{\delta H}{\delta \shift^i}=-\tetrad^A{}_i \partial_j \momenta_A{}^j-\momenta_A{}^j T^A{}_{ji},
	\end{split}
\end{align}
is independent of the teleparallel theory, though it is slightly simpler than the momenta constraint of the covariant formulation. 
The time evolution of the conjugate momenta with respect to the spatial tetrads is quite lengthy:
\begin{align}
\label{eq:dpiGauge}
	\begin{split}
	-\dot{\momenta}_A{}^i&=\frac{\delta H}{\delta \tetrad^A{}_i}=\lapse \left(\frac{\sqrt{\inducedmetric}}{2\kappa}\tetrad_A{}^i\left(V(\phi)-{}^3 \mathbb{T} \right)  -\frac{\sqrt{\inducedmetric}}{\kappa}T^C{}_{mn}T^B{}_{kl}\left(\tetrad_A{}^m H_{CB}{}^{inlk}+\tetrad_A{}^k H_{CB}{}^{mnli}\right. \right. \\
	&\left. \left. -\frac{1}{2}\normalvector_A\normalvector_C\tetrad_B{}^{[m}\inducedmetric^{n][k}\inducedmetric^{l]i}+ \normalvector_A \normalvector_B\tetrad_C{}^{[m}\inducedmetric^{n][k}\inducedmetric^{l]i} \right)+\inducedmetric^{im}\normalvector_A\tetrad^C{}_m\partial_j \momenta_C{}^j \right) \\
	& -\shift^i\partial_j\momenta_A{}^j+\partial_j \left(\frac{2\lapse\sqrt{\inducedmetric}}{\kappa}H_{BA}{}^{kl[ij]}T^B{}_{kl}+2\shift^{[i}\momenta_A{}^{j]}+2{}^{\mathcal{V}}\lambda^{[i} \tetrad_A{}^{j]}-{}^{\mathcal{A}}\lambda_{[kl]} \inducedmetric^{k[i}\inducedmetric^{j]l}\normalvector_A \right)\\
	&+\frac{\lapse\kappa}{2\sqrt{\inducedmetric}\phi} \left(\momenta_C{}^{(i}\momenta_B{}^{l)}\tetrad^C{}_j\tetrad^B{}_k\eta_{AD}\inducedmetric^{jk}\tetrad^D{}_l-\momenta_C{}^m\momenta_B{}^l\tetrad^C{}_j\tetrad^B{}_k\tetrad_A{}^{(j}\inducedmetric^{k)i}\inducedmetric_{ml}+2\momenta_A{}^j\momenta_B{}^{(k}\inducedmetric^{i)l}\inducedmetric_{jk}\tetrad^B{}_l  \right. \\
 &\left. -\momenta_A{}^i\momenta_B{}^j\tetrad^B{}_j-\tetrad_A{}^i\momenta_C{}^m\momenta_B{}^{(l}\inducedmetric^{k)j}\tetrad^C{}_k\tetrad^B{}_j\inducedmetric_{ml}+\frac{1}{2}\tetrad_A{}^i\momenta_C{}^k\momenta_B{}^j\tetrad^C{}_k\tetrad^B{}_j\right)  \\
	&+{}^{\mathcal{V}}\lambda_j \left(-\frac{\kappa}{\sqrt{\inducedmetric}\phi}\inducedmetric^{im}\normalvector_A\tetrad^C{}_m\momenta_C{}^j -\frac{\kappa}{\sqrt{\inducedmetric}\phi}\normalvector^B\momenta_B{}^j\tetrad_A{}^i +T^B{}_{kj}\left(\tetrad_A{}^k \tetrad_B{}^i+\normalvector_A\normalvector_B \inducedmetric^{ik}\right)\right. \\
	& \left.+2 T^B{}_{kl}\tetrad_B{}^k\tetrad_A{}^{(j}\inducedmetric^{l)i}
	\right)+{}^{\mathcal{A}}\lambda_{[jk]} \left(c_2\normalvector_A\tetrad_B{}^i\inducedmetric^{jn}\inducedmetric^{km}T^B{}_{nm}+\frac{\kappa}{\sqrt{\inducedmetric}\phi}\inducedmetric^{jl}\momenta_B{}^k\tetrad^B{}_l\tetrad_A{}^i \right. \\
	&\left. - \inducedmetric^{i[k}\inducedmetric^{j]m}\normalvector_B T^B{}_{ml}\tetrad_A{}^l- \inducedmetric^{im}\normalvector_B T^B{}_{nm}\tetrad_A{}^{[k}\inducedmetric{}^{j]n}+\frac{2\kappa}{\sqrt{\inducedmetric}\phi} \momenta_B{}^k \tetrad^B{}_l\tetrad_A{}^{[l}\inducedmetric^{j]i}+\frac{\kappa}{\sqrt{\inducedmetric}\phi}\inducedmetric^{ik}\momenta_A{}^j \right)
	\end{split}
	\end{align}
and the time evolution of the momenta with respect to the scalar field reads
	\begin{align}
	\begin{split}
	-{}^{\phi}\dot{\momenta}&=\frac{\delta H}{\delta \phi}=\lapse\left[  -\frac{\kappa }{2\tetrad \phi^2}\left(\inducedmetric^{kl}\inducedmetric_{i(j} \tetrad^A{}_{k)} \tetrad^B{}_l\momenta_A{}^i \momenta_B{}^j  -\frac{\momenta_A{}^i\momenta_B{}^j\tetrad^A{}_i \tetrad^B{}_j }{2}\right) +\frac{ \sqrt{\inducedmetric}}{2\kappa}\frac{\delta V(\phi)}{\delta \phi} -\frac{\sqrt{\inducedmetric}}{2\kappa\phi}{}^3 \mathbb{T}\right]\\
	&-{}^\mathcal{V}\lambda_i \frac{\kappa \normalvector^A \momenta_A{}^i }{\tetrad \phi^2}+{}^\mathcal{A}\lambda_{[ik]}\frac{\kappa \inducedmetric^{ij}\momenta_A{}^k \tetrad^A{}_j}{\tetrad \phi^2},
	\end{split}
	\end{align}
which is similar to the case of $f(R)$ gravity (see Eq. (3.10) in \cite{Deruelle:2009zk}).
The time evolution of the lapse function and shift vector are again given by Lagrange multipliers:
	\begin{align}
	\begin{split}
	\dot{\lapse}=\frac{\delta H}{\delta {}^{\lapse}\momenta}=-{}^{\lapse}\lambda,
	\end{split}
	\end{align}
	
	\begin{align}
	\dot{\beta}^i=\frac{\delta H}{\delta {}^\beta\pi_i}=-{}^\shift \lambda^i.
	\end{align}
 The time evolution of the spatial tetrad is given by
	\begin{align}
	\begin{split}
	\dot{\theta}^A{}_i=\frac{\delta H}{\delta \pi_A{}^i}&= \lapse\left[
	\frac{ \kappa}{\sqrt{\inducedmetric} \phi }\left(\inducedmetric^{kl}\inducedmetric_{i(j}\tetrad^A{}_{k)}\tetrad^B{}_l \momenta_B{}^j -\frac{\momenta_B{}^j\tetrad^A{}_i \tetrad^B{}_j }{2}\right)+\partial_i \normalvector^A \right]-\shift^jT^A{}_{ij}\\
	&+{}^\mathcal{V}\lambda_i \frac{\kappa \normalvector^A}{\sqrt{\inducedmetric} \phi}+{}^\mathcal{A}\lambda_{[ik]}\frac{\kappa\inducedmetric^{jk}\tetrad^A{}_j}{\sqrt{\inducedmetric} \phi}+\partial_i\left(\shift^j \tetrad^A{}_j \right).
	\end{split}
	\end{align}
Finally, the time evolution of the scalar field is also given by a Lagrange multiplier and reads as:
	
	\begin{align}
	\dot{\phi}=\frac{\delta H}{\delta {}^\phi \pi}=-{}^\phi \lambda. 
	\end{align}
 As we can see from the above relations, in the Weitzenböck gauge the set of equations and primary constraints are fewer. Also note that many equations in the covariant formulation implicitly depend on the spin connection through torsion. Therefore, Hamilton's equations simplify even more than it appears by a first look in the Weitzenböck gauge. Thus, for calculation purposes, this formulation has a great advantage. Considering that the computation of Poisson brackets has given contradicting results \cite{Li:2011rn,Blagojevic:2020dyq,Ferraro:2018tpu}, it is evident that the lengthiness of such calculation increases the probability of committing mistakes. The arguments reported in \cite{Golovnev:2022rui,Blixt:2022rpl} show that both formulations are equally valid and, moreover, in \cite{Golovnev:2023yla} the authors argue that this formulation is even more fundamental. That being said, the covariant formulation may still have advantages (see the previous subsection for some discussion on the topic), which is why both formulations are presented in this article.

	\section{The Hamilton equations for New General Relativity}
\label{sec:HeqNGR}
	The one-parameter family of NGR is given by the choice $c_1=-\dfrac{1}{4}(\rho+1)$, $c_2=\dfrac{1}{2}(\rho-1)$ and $c_3=1$. In this case, $\mathcal{B}_\mathcal{V}=0$ and only ${}^{\mathcal{V}}\lambda_i$ is non-vanishing. Furthermore, $\phi=1$, while ${}^\mathcal{A}\lambda_{ij}$, ${}^\mathcal{S}\lambda_{ij}$, ${}^\mathcal{T}\lambda$ and ${}^\mathcal{\phi}\lambda$ are all absent (from the $f(T_\mathrm{NGR})$ point of view they are zero). TEGR is recovered in the limit $\rho =0$, which also implies that $\mathcal{B}_\mathcal{A}=0$ and ${}^\mathcal{A}\lambda_{ij} \neq 0$. Furthermore, for the one-parameter family of NGR, it is $\mathcal{B}_\mathcal{A}=\frac{1}{2c_1-c_2}=-\frac{1}{\rho}$, $\mathcal{B}_\mathcal{S}=\frac{1}{2c_1+c_2	}=-1$, and $\mathcal{B}_\mathcal{T}=\frac{1}{2c_1+c_2+3c_3}=\dfrac{1}{2}$. The theory includes primary constraints that do not belong to the first-class category \cite{Cheng:1988zg,Tomonari:2024ybs} (refer to Eq. (56) in \cite{Tomonari:2024ybs}). Given that Hamilton's equations hold exclusively on the constraint surface \cite{Bojowald_2010}, one needs to include all constraints during the analysis, checking their evaluation over time.
		
	\subsection{Without gauge fixing}
\label{sec:HeqNGRw}
We start by presenting the covariant Hamiltonian for NGR, which reads as: 
	\begin{align}
	\begin{split}
	\mathcal{H}_{\mathrm{NGR}}&=\lapse  \left[ \frac{\sqrt{\inducedmetric}}{2\rho\kappa} {}^{\mathcal{A}}C^{ij} {}^{\mathcal{A}} C_{ij}+\frac{\sqrt{\inducedmetric}}{2\kappa} {}^{\mathcal{S}}C^{ij} {}^{\mathcal{S}} C_{ij}-\frac{3\sqrt{\inducedmetric}}{4\kappa} {}^{\mathcal{T}}C {}^{\mathcal{T}} C-\frac{\sqrt{\inducedmetric}}{2\kappa}{}^3 \mathbb{T} -\normalvector^A\partial_i \momenta_A{}^i+\momenta_A{}^i\spinconnection^A{}_{Bi}\normalvector^B \right]\\ & +\shift^j\left[ -\tetrad^A{}_j \partial_i \momenta_A{}^i+\momenta_A{}^i\spinconnection^A{}_{Ci} \tetrad^C{}_j-\momenta_A{}^i T^A{}_{ij} \right] -{}^\alpha \lambda {}^\alpha \pi-{}^\beta \lambda_i {}^\beta \pi^i\\
 & -\lambda_{AB}\left( P^{[A}{}_D\eta^ {B]C}\Lorentz_C{}^D+\momenta_C{}^i\eta^{C[B}\tetrad^{A]}{}_i \right) -{}^{\mathcal{V}}\lambda_i\left(\frac{{}^{\mathcal{V}} \pi^i\kappa}{\sqrt{\inducedmetric}}+ T^B{}_{jk}\inducedmetric^{ik}\inducedmetric^{jl}\tetrad^A{}_l\eta_{AB}\right).
	\end{split}
	\end{align}
Compared to TEGR, the Hamiltonian constraint contains extra terms with the antisymmetric part of the conjugate momenta:
	
		\begin{align}
	\begin{split}
	-{}^\alpha \dot{\pi}=\frac{\delta H}{\delta \alpha}&=\frac{\sqrt{\inducedmetric}}{2\rho\kappa} {}^{\mathcal{A}}C^{ij} {}^{\mathcal{A}} C_{ij}+\frac{\sqrt{\inducedmetric}}{2\kappa} {}^{\mathcal{S}}C^{ij} {}^{\mathcal{S}} C_{ij}-\frac{3\sqrt{\inducedmetric}}{4\kappa} {}^{\mathcal{T}}C {}^{\mathcal{T}} C  -\frac{\sqrt{\inducedmetric}}{2\kappa}{}^3 \mathbb{T}-\normalvector^A\partial_i \momenta_A{}^i+\momenta_A{}^i\spinconnection^A{}_{Bi}\normalvector^B, 
	\end{split}
	\end{align}
	while the momenta constraint remains unchanged
	\begin{align}
	-{}^\beta \dot{\pi}_i=\frac{\delta H}{\delta \beta^i}=-\tetrad^A{}_i \partial_j \momenta_A{}^j+\momenta_A{}^j\spinconnection^A{}_{Cj} \tetrad^C{}_i-\momenta_A{}^j T^A{}_{ji} .
	\end{align}

	The explicit expression for the time evolution of the conjugate momenta is very lengthy and for this reason it is dissected into smaller pieces, with their explicit expressions given by   Eqs. \eqref{eq:NGRlapse}-\eqref{eq:NGRLV} 
	\begin{align}
	\begin{split}
	-\dot{\momenta}_A{}^i&=\frac{\delta H}{\delta \tetrad^A{}_i}=-\lapse {}^{\lapse}\dot{\momenta}_A{}^i-\shift^j {}^{\shift}\dot{\momenta}_A{}^{i}{}_j-\partial_j {}^{\partial} \dot{\momenta}_A{}^{ij}-\lambda_{[BC]}{}^{\omega}\dot{\momenta}_A{}^{iBC}+\frac{{}^{\mathcal{BA}}\dot{\momenta}_A{}^i}{\rho}\\
	&+{}^{\mathcal{BS}}\dot{\momenta}_A{}^i -\frac{1}{2}{}^{\mathcal{BT}}\dot{\momenta}_A{}^i  -{}^{\mathcal{V}}\lambda_j {}^{\lambda\mathcal{V}}\dot{\momenta}_A{}^{ij}.
	\end{split}
	\end{align}
Similarly, the explicit time evolution of the conjugate momenta with respect to the Lorentz matrices, is presented in \eqref{eq:NGRlapse}-\eqref{eq:NGRLMomLV}, while below we only report its implicit expression:
	\begin{align}
	\begin{split}
	-\dot{P}^A{}_B&=\frac{\delta H}{\delta \Lambda_A{}^B}=-\lapse {}^{\lapse}\dot{P}^A{}_B-\shift_i{}^{\shift}\dot{P}^A{}_B{}^i-\partial_i {}^{\partial}\dot{P}^A{}_B{}^i-\lambda_{[CD]}{}^{\omega}\dot{P}^A{}_B{}^{CD}+\frac{{}^{\mathcal{BA}}\dot{P}^A{}_B}{\rho}-{}^{\mathcal{V}}\lambda_i {}^{\lambda\mathcal{V}}\dot{P}^A{}_B{}^i
	\end{split}
	\end{align}
	As expected from diffeomorphism invariance, the time evolution of lapse $\lapse$ and shift $\shift^i$ are  proportional to the Lagrange multipliers:
	\begin{align}
	\dot{\alpha}=\frac{\delta H}{\delta {}^\alpha \pi}=-{}^\lapse \lambda, 
	\end{align}
	\begin{align}
	\dot{\beta}^i=\frac{\delta H}{\delta {}^\beta\pi_i}=-{}^\shift \lambda^i.
	\end{align}
 The time evolution of the tetrad gets a couple of extra terms with respect to TEGR, while the part related to ${}^\mathcal{A}\lambda_{ij}$ is absent:
	\begin{align}
	\begin{split}
	\dot{\theta}^A{}_i=\frac{\delta H}{\delta \pi_A{}^i}&= \lapse\left[ \left(\frac{\kappa\inducedmetric^{kl}\inducedmetric_{i[j}\tetrad^A{}_{k]}\tetrad^B{}_l \momenta_B{}^j}{\rho\sqrt{\inducedmetric}  }-\frac{\rho-1}{2\rho} \inducedmetric^{jk}\normalvector_B T^B{}_{ij}\tetrad^A{}_k \right)\right. \\
	&\left. +\frac{ \kappa}{\sqrt{\inducedmetric} }\left(\inducedmetric^{kl}\inducedmetric_{i(j}\tetrad^A{}_{k)}\tetrad^B{}_l \momenta_B{}^j -\frac{2\momenta_B{}^j\tetrad^A{}_i \tetrad^B{}_j }{3}\right)+\normalvector^B\omega^A{}_{Bi} +\partial_i\normalvector^A \right]\\
	&+\shift^j\left[\tetrad^B{}_j \omega^A{}_{Bi}-T^A{}_{ij}\right]+\lambda_{[CB]}\tetrad^B{}_i\eta^{AC}+{}^\mathcal{V}\lambda_i \frac{\kappa \normalvector^A}{\sqrt{\inducedmetric} }.
	\end{split}
	\end{align}
 The time evolution of the Lorentz matrix is only related to terms occurring in the corresponding primary constraint, which are the same for all teleparallel theories:
	\begin{align}
	\begin{split}
	\dot{\Lambda}_A{}^B&=\frac{\delta H}{\delta P^A{}_B}= \lambda_{[DA]}\Lambda_C{}^B \eta^{CD}\,.
	\end{split}
	\end{align}
	
	Below, we list the explicit expressions for $\dot{\momenta}_A{}^i$, starting with
	\begin{align}
	\begin{split}
	-{}^{\lapse}\dot{\momenta}_A{}^i&=-\frac{2\sqrt{\inducedmetric}}{\kappa}H_{CB}{}^{mn[ki]}T^C{}_{mn}\omega^B{}_{Ak}-\frac{\sqrt{\inducedmetric}}{2\kappa}\tetrad_A{}^i{}^3 \mathbb{T}  -\frac{\sqrt{\inducedmetric}}{\kappa}T^C{}_{mn}T^B{}_{kl}\left(\tetrad_A{}^m H_{CB}{}^{inlk}+\tetrad_A{}^k H_{CB}{}^{mnli}\right. \\
	&\left. +\frac{\rho-1}{2}\normalvector_A\normalvector_C\tetrad_B{}^{[m}\inducedmetric^{n][k}\inducedmetric^{l]i}+\normalvector_A \normalvector_B\tetrad_C{}^{[m}\inducedmetric^{n][k}\inducedmetric^{l]i} \right)+\inducedmetric^{im}\normalvector_A\tetrad^C{}_m\left(\partial_j \momenta_C{}^j-\momenta_B{}^j\omega^B{}_{Cj}\right),
	\end{split}
	\end{align}
	where it is worth noticing that the constant $\rho$ is included in the super-metric $H_{AB}{}^{ijkl}(\rho)$. The term related to the shift is, as expected, the same as other teleparallel theories
	\begin{align}
 \label{eq:NGRlapse}
	\begin{split}
	-{}^{\shift}\dot{\momenta}_A{}^{i}{}_j&= \momenta_B{}^{i}\omega^B{}_{Aj}-\delta_j^i\partial_k\momenta_A{}^k.
	\end{split}
	\end{align}
	Similarly, in the computation of ${}^{\partial}\dot{\momenta}_A{}^{ij}$, the parameter $\rho$ appears below through the super-metric $H_{BA}{}^{klij}(\rho)$ and also differs due to the different set of primary constraints
	\begin{align}
 \label{eq:NGRshift}
	\begin{split}
	-{}^{\partial}\dot{\momenta}_A{}^{ij}&=\frac{2\lapse\sqrt{\inducedmetric}}{\kappa}H_{BA}{}^{kl[ij]}T^B{}_{kl}+2\shift^{[i}\momenta_A{}^{j]}+2{}^{\mathcal{V}}\lambda^{[i} \tetrad_A{}^{j]}.
	\end{split}
	\end{align}
	The next part of $\dot{\momenta}_A{}^i$ comes from the primary constraints associated with the covariant formulation. Since these primary constraints are independent of the given theory, they clearly coincide with those of TEGR and $f(T)$ gravity and yield:
	
	\begin{align}
 \label{eq:NGRomega}
	\begin{split}
	-{}^{\omega}\dot{\momenta}_A{}^{iBC}&=\delta^{[C}_A \eta^{B]D}\momenta_D{}^i.
	\end{split}
	\end{align}
	In both TEGR and $f(T)$ gravity, it is $\mathcal{B}_\mathcal{A}=0$, whereas in NGR, it is  $\mathcal{B}_\mathcal{A}=-\dfrac{1}{\rho}$ and this non-vanishing expression gives rise to extra terms in the Hamiltonian. The contribution of $\mathcal{B}_\mathcal{A}$ to $\dot{\momenta}_A{}^i$ is presented below:

	\begin{align}
 \label{eq:NGRBA}
	\begin{split}
	-{}^{\mathcal{BA}}\dot{\momenta}_A{}^i&= \lapse (\rho-1) \inducedmetric^{n[i}\momenta_B{}^{k]}\normalvector_C \tetrad^B{}_n\omega^C{}_{Ak}-\frac{\lapse (\rho-1)^2 \sqrt{\inducedmetric}}{2\kappa}\inducedmetric^{n[k}\inducedmetric^{i]j}\normalvector_C \normalvector_B T^B{}_{nj}\omega^C{}_{Ak}\\
	&+\partial_k \left[ \lapse (\rho-1) \inducedmetric^{n[k}\momenta_B{}^{i]}\normalvector_A \tetrad^B{}_n-\frac{\lapse (\rho-1)^2 \sqrt{\inducedmetric}}{2\kappa}\inducedmetric^{n[i}\inducedmetric^{k]j}\normalvector_A \normalvector_B T^B{}_{nj} \right]\\
	&+\lapse \inducedmetric^{im}\eta_{CD}\normalvector_A \tetrad^D{}_m\inducedmetric^{kl}T^C{}_{kj}\left(\frac{\rho-1}{2} \momenta_B{}^j \tetrad^B{}_l+\frac{(\rho-1)^2}{4\kappa}\sqrt{\inducedmetric}\inducedmetric^{jn}\normalvector_B T^B{}_{ln} \right)\\
	&+\frac{\lapse \kappa}{2\sqrt{\inducedmetric}} \inducedmetric_{jk}\inducedmetric^{l[n}\momenta_B{}^{j]}\momenta_D{}^k \tetrad^B{}_l \tetrad_A{}^i\tetrad^D{}_n-\frac{\lapse (\rho-1)^2 \sqrt{\inducedmetric}}{8\kappa}\inducedmetric^{jl}\inducedmetric^{kn}\normalvector_B\normalvector_D T^B{}_{jk}T^D{}_{ln}\tetrad_A{}^i\\
	&+\lapse (\rho-1) \normalvector_B \momenta_C{}^j T^B{}_{lj}\tetrad^C{}_m\tetrad_A{}^{(l}\inducedmetric^{m)i}+\frac{\lapse\kappa}{2\sqrt{\inducedmetric}}\inducedmetric^{in}\inducedmetric_{jl}\momenta_B{}^j \momenta_C{}^l \tetrad^B{}_k \tetrad^C{}_n \tetrad_A{}^k\\
	&-\frac{\lapse\kappa}{2\sqrt{\inducedmetric}} \inducedmetric^{jk}\eta_{AD}\momenta_B{}^i\momenta_C{}^l\tetrad^B{}_j\tetrad^C{}_k\tetrad^D{}_l+\frac{2\lapse\sqrt{\inducedmetric}}{\kappa}\inducedmetric^{in}\inducedmetric^{l[k}\inducedmetric^{m]j}\eta_{AD}\normalvector_B \normalvector_D T^B{}_{jl}T^C{}_{mn}\tetrad^D{}_k \\
	&+ \frac{\lapse(\rho-1)}{2}\inducedmetric^{ik}\normalvector_B\momenta_A{}^j T^B{}_{kj}+\frac{\lapse \kappa}{\sqrt{\inducedmetric}}\momenta_B{}^{[i}\inducedmetric^{k]l}\inducedmetric_{jk}\momenta_A{}^j\tetrad^B{}_l.
	\end{split}
	\end{align}
	Interestingly, though Eq. \eqref{eq:NGRBA} is a very lengthy expression, it greatly simplifies for the particular case $\rho=1$.  The next two parts of $\dot{\momenta}_A{}^i$, namely ${}^{\mathcal{BS}}\dot{\momenta}_A{}^i$ and ${}^{\mathcal{BT}}\dot{\momenta}_A{}^i$, appear also in TEGR and can be combined in the following way to simplify the overall expression:
 \begin{align}
  \label{eq:NGRBT}
     \begin{split}
         {}^{\mathcal{BS}}\dot{\momenta}_A{}^i-\frac{1}{2}{}^{\mathcal{BT}}\dot{\momenta}_A{}^i&=\frac{\lapse\kappa}{2\sqrt{\inducedmetric}} \left(\momenta_C{}^{(i}\momenta_B{}^{l)}\tetrad^C{}_j\tetrad^B{}_k\eta_{AD}\inducedmetric^{jk}\tetrad^D{}_l-\momenta_C{}^m\momenta_B{}^l\tetrad^C{}_j\tetrad^B{}_k\tetrad_A{}^{(j}\inducedmetric^{k)i}\inducedmetric_{ml}\right. \\
 & \left. +2\momenta_A{}^j\momenta_B{}^{(k}\inducedmetric^{i)l}\inducedmetric_{jk}\tetrad^B{}_l   -\momenta_A{}^i\momenta_B{}^j\tetrad^B{}_j-\tetrad_A{}^i\momenta_C{}^m\momenta_B{}^{(l}\inducedmetric^{k)j}\tetrad^C{}_k\tetrad^B{}_j\inducedmetric_{ml} \right. \\
 & \left. +\frac{1}{2}\tetrad_A{}^i\momenta_C{}^k\momenta_B{}^j\tetrad^C{}_k\tetrad^B{}_j \right).
     \end{split}
 \end{align}

	
	Also the terms proportional to ${}^\mathcal{V}\lambda_j$ coincide with those of TEGR:
	\begin{align}
 \label{eq:NGRLV}
	\begin{split}
	-{}^{\lambda\mathcal{V}}\dot{\momenta}_A{}^{ij}&=-2\inducedmetric^{j[i}\tetrad_B{}^{k]}\omega^B{}_{Ak}-\frac{\kappa}{\sqrt{\inducedmetric}}\inducedmetric^{im}\normalvector_A\tetrad^C{}_m\momenta_C{}^j\\
	&-\frac{\kappa}{\sqrt{\inducedmetric}}\normalvector^B\momenta_B{}^j\tetrad_A{}^i+T^B{}_{kj}\left(\tetrad_A{}^k \tetrad_B{}^i+\normalvector_A\normalvector_B \inducedmetric^{ik}\right)+2 T^B{}_{kl}\tetrad_B{}^k\tetrad_A{}^{(j}\inducedmetric^{l)i},
	\end{split}
	\end{align}
 while there are no terms associated with ${}^\mathcal{A}\lambda_{jk}$ in NGR. 

 Considering now the explicit expression for the constituents of $\dot{P}^A{}_B$, we have

	\begin{align}
 \label{eq:NGRLMomlapse}
	\begin{split}
	- {}^{\lapse}\dot{P}^A{}_B&=- \left( \Lambda^{-1}\right)^A{}_D\omega^C{}_{Bi}\normalvector^D\momenta_C{}^i -(H_{CE}{}^{kijl}-H_{CE}{}^{iljk})\frac{ \sqrt{\inducedmetric} \left(\Lambda^{-1} \right)^A{}_D \omega^C{}_{Bi}  T^E{}_{jl}\tetrad^D{}_k}{\kappa},
	\end{split}
	\end{align}
	depending implicitly on $\rho$ through the supermetric $H_{AB}{}^{ijkl}(\rho)$. The part related to shift, namely
	\begin{align}
 \label{eq:NGRLMomshift}
	\begin{split}
	-{}^{\shift}\dot{P}^A{}_B{}^i&= -\momenta_C{}^{j} \left(\Lambda^{-1} \right)^A{}_D \omega^C{}_{Bi} \tetrad^D{}_j,
	\end{split}
	\end{align}
	 is theory independent. The next part, namely ${}^{\partial}\dot{P}^A{}_B{}^i$, again depends implicitly on $\rho$ through the super metric. In addition, the different set of primary constraints alters its expression with respect to TEGR:
	\begin{align}
 \label{eq:NGRLder}
	\begin{split}
	- {}^{\partial}\dot{P}^A{}_B{}^i&=\partial_i \left[\lapse \left(\Lambda^{-1} \right)^A{}_D \normalvector^D \pi_B{}^i-\shift^i \left(\Lambda^{-1} \right)^A{}_D \momenta_B{}^k \right. \\
	& \left. +\frac{\lapse}{\kappa}\left(H_{BC}{}^{kijl}-H_{BC}{}^{iljk} \right)\sqrt{\inducedmetric} \left(\Lambda^{-1} \right)^A{}_D T^C{}_{jl} \tetrad^D{}_k   +2{}^\mathcal{V}\lambda_l  \left(\Lambda^{-1} \right)^A{}_D \inducedmetric^{i[j}\inducedmetric^{l]k}\eta_{BC} \tetrad^C{}_j\tetrad^D{}_k \right].
	\end{split}
	\end{align}
	The term related to the covariant primary constraint is theory independent and reads:
	\begin{align}
 \label{eq:NGRLMomomega}
	\begin{split}
	-{}^{\omega}\dot{P}^A{}_B{}^{CD}&=\eta^{A[C}P^{D]}{}_B.
	\end{split}
	\end{align}
	As next, there is another part which does not occur in TEGR due to the condition $\mathcal{B}_\mathcal{A}=0$. It is:

	\begin{align}
 \label{eq:NGRLMomBA}
	\begin{split}
	-{}^{\mathcal{BA}}\dot{P}^A{}_B&=-\lapse (\rho-1) \left(\Lambda^{-1} \right)^A{}_D \omega^C{}_{Bi} \inducedmetric^{j[k}\momenta_E{}^{i]}\normalvector_C  \tetrad^D{}_k\tetrad^E{}_j\\
	&-\frac{\lapse (\rho-1)^2 \sqrt{\inducedmetric}}{2\kappa}  \left(\Lambda^{-1} \right)^A{}_D \omega^C{}_{Bi} \inducedmetric^{i[l}\inducedmetric^{j]k}\normalvector_C \normalvector_E  T^E{}_{jl}\tetrad^D{}_k \\
	&+\partial_i \left[\lapse (\rho-1) \left(\Lambda^{-1} \right)^A{}_D \inducedmetric^{j[k}\momenta_C{}^{i]} \normalvector_B \tetrad^C{}_j \tetrad^D{}_k+\frac{\lapse (\rho-1)^2 \sqrt{\inducedmetric}}{2\kappa} \left(\Lambda^{-1} \right)^A{}_D \inducedmetric^{i[l}\inducedmetric^{j]k}\normalvector_B\normalvector_C T^C{}_{jl}\tetrad^D{}_k   \right].
	\end{split}
	\end{align}
	Again, the case $\rho=1$ simplifies, or in this case completely trivialize, the expression. Other values of $\rho$ significantly extend the already very cumbersome calculation of the time evolution of primary constraints in NGR. However, note that this term only plays a role in the time evolution of the primary constraints associated to the covariant formulation ${}^\omega \dot{C}_{AB}$. Lastly, there is only one more term, which is related to the Lagrange multiplier ${}^\mathcal{V}\lambda_j$:
	\begin{align}
 \label{eq:NGRLMomLV}
	\begin{split}
	- {}^{\lambda\mathcal{V}}\dot{P}^A{}_B{}^i&=2  \left(\Lambda^{-1} \right)^A{}_D \omega^C{}_{Bl} \tetrad_C{}^{[i}\inducedmetric^{l]k}  \tetrad^D{}_k . 
	\end{split}
	\end{align}
These are the Hamilton equations for NGR and, together with the primary constraints, they determine the evolution of the fields. In order to recover TEGR, one needs either to set $\rho=0$ or to remove the terms containing $\frac{1}{\rho}$. In addition, the primary constraint ${}^\mathcal{A}C_{ij}$ needs to be considered. For this reason it is slightly easier to compare $f(T)$ with TEGR, rather than NGR, since $f(T)$ has the same number of primary constraints, aside the occurrence of the scalar field $\phi$. The time evolution of the vector constraints ${}^{\mathcal{V}}\dot{C}_i$ for NGR has only been investigated in \cite{Cheng:1988zg}. To conclude, we again point out that the covariant formulation evidently makes these cumbersome calculations even more lengthy.

		\subsection{The Weitzenböck gauge}
  \label{sec:HeqNGRw0}
In the Weitzenböck gauge, the Hamiltonian becomes
	\begin{align}
	\begin{split}
	\mathcal{H}_{\mathrm{NGR}}&=\lapse  \left[ \frac{\sqrt{\inducedmetric}}{2\rho\kappa} {}^{\mathcal{A}}C^{ij} {}^{\mathcal{A}} C_{ij}+\frac{\sqrt{\inducedmetric}}{2\kappa} {}^{\mathcal{S}}C^{ij} {}^{\mathcal{S}} C_{ij}-\frac{3\sqrt{\inducedmetric}}{4\kappa} {}^{\mathcal{T}}C {}^{\mathcal{T}} C-\frac{\sqrt{\inducedmetric}}{2\kappa}{}^3 \mathbb{T}-\normalvector^A\partial_i \momenta_A{}^i\right] \\
 &+\shift^j\left[ -\tetrad^A{}_j \partial_i \momenta_A{}^i-\momenta_A{}^i T^A{}_{ij} \right] -{}^\alpha \lambda {}^\alpha \pi-{}^\beta \lambda_i {}^\beta \pi^i -{}^{\mathcal{V}}\lambda_i\left(\frac{{}^{\mathcal{V}} \pi^i\kappa}{\sqrt{\inducedmetric}}+ T^B{}_{jk}\inducedmetric^{ik}\inducedmetric^{jl}\tetrad^A{}_l\eta_{AB}\right).
	\end{split}
	\end{align}

	Below, we present the Hamilton equations starting from the Hamiltonian constraint
		\begin{align}
	\begin{split}
	-{}^\alpha \dot{\pi}=\frac{\delta H}{\delta \alpha}&=\frac{\sqrt{\inducedmetric}}{2\rho\kappa} {}^{\mathcal{A}}C^{ij} {}^{\mathcal{A}} C_{ij}+\frac{\sqrt{\inducedmetric}}{2\kappa} {}^{\mathcal{S}}C^{ij} {}^{\mathcal{S}} C_{ij}-\frac{3\sqrt{\inducedmetric}}{4\kappa} {}^{\mathcal{T}}C {}^{\mathcal{T}} C  -\frac{\sqrt{\inducedmetric}}{2\kappa}{}^3 \mathbb{T}-\normalvector^A\partial_i \momenta_A{}^i ,
	\end{split}
	\end{align}
differing significantly to TEGR due to the lack of the primary constraint ${}^\mathcal{A}C_{ij}\approx 0$, which implies the occurrence of the first term in the above expression. The momenta constraint, on the other hand, remains the same for all teleparallel theories. It is
 
	\begin{align}
	-{}^\beta \dot{\pi}_i=\frac{\delta H}{\delta \beta^i}=-\tetrad^A{}_i \partial_j \momenta_A{}^j-\momenta_A{}^j T^A{}_{ji} .
	\end{align}
 The lengthiest expression is the time evolution of the conjugate momenta $\dot{\momenta}_A{}^i$. It is even more lengthy in NGR than $f(T)$ gravity (see Eq. \eqref{eq:dpiGauge}). For this reason we divide it in the following way

	\begin{align}
	\begin{split}
	-\dot{\momenta}_A{}^i&=\frac{\delta H}{\delta \tetrad^A{}_i}=-\lapse {}^{\lapse}\dot{\momenta}_A{}^i-\shift^j {}^{\shift}\dot{\momenta}_A{}^{i}{}_j-\partial_j {}^{\partial} \dot{\momenta}_A{}^{ij}+\frac{{}^{\mathcal{BA}}\dot{\momenta}_A{}^i}{\rho}+{}^{\mathcal{BS}}\dot{\momenta}_A{}^i -\frac{1}{2}{}^{\mathcal{BT}}\dot{\momenta}_A{}^i  -{}^{\mathcal{V}}\lambda_j {}^{\lambda\mathcal{V}}\dot{\momenta}_A{}^{ij},
	\end{split}
	\end{align}
whose explicit expression are outlined in Eq. \eqref{eq:GaugeNGRlapse}-\eqref{eq:GaugeNGRLV}. The time evolution of lapse $\dot{\lapse}$ and $\dot{\shift}^i$ are, as expected from diffeomorphism invariance, proportional to Lagrange multipliers, \emph{i.e.}

	\begin{align}
	\dot{\alpha}=\frac{\delta H}{\delta {}^\alpha \pi}=-{}^\lapse \lambda, 
	\end{align}
	\begin{align}
	\dot{\beta}^i=\frac{\delta H}{\delta {}^\beta\pi_i}=-{}^\shift \lambda^i.
	\end{align}
 The evolution of the spatial tetrads explicitly contains the parameter $\rho$ which, however, does not appear implicitly in ${}^\mathcal{A}\lambda_{ij}$, unlike TEGR:
	\begin{align}
	\begin{split}
	\dot{\theta}^A{}_i=\frac{\delta H}{\delta \pi_A{}^i}&= \lapse\left[ \left(\frac{\kappa\inducedmetric^{kl}\inducedmetric_{i[j}\tetrad^A{}_{k]}\tetrad^B{}_l \momenta_B{}^j}{\rho\sqrt{\inducedmetric}  }-\frac{\rho-1}{2\rho} \inducedmetric^{jk}\normalvector_B T^B{}_{ij}\tetrad^A{}_k \right)\right. \\
	&\left. +\frac{ \kappa}{\sqrt{\inducedmetric}  }\left(\inducedmetric^{kl}\inducedmetric_{i(j}\tetrad^A{}_{k)}\tetrad^B{}_l \momenta_B{}^j -\frac{2\momenta_B{}^j\tetrad^A{}_i \tetrad^B{}_j }{3}\right)+\partial_i\normalvector^A \right]-\shift^jT^A{}_{ij}+{}^\mathcal{V}\lambda_i \frac{\kappa \normalvector^A}{\sqrt{\inducedmetric} }.
	\end{split}
	\end{align}
	
	Below, we write the explicit expressions of $\dot{\momenta}_A{}^i$, by listing all the terms contained in its definition. Let us first consider ${}^{\lapse}\dot{\momenta}_A{}^i$, whose form is given by:
	\begin{align}
 \label{eq:GaugeNGRlapse}
	\begin{split}
	-{}^{\lapse}\dot{\momenta}_A{}^i&=-\frac{\sqrt{\inducedmetric}}{2\kappa}\tetrad_A{}^i{}^3 \mathbb{T}  -\frac{\sqrt{\inducedmetric}}{\kappa}T^C{}_{mn}T^B{}_{kl}\left(\tetrad_A{}^m H_{CB}{}^{inlk}+\tetrad_A{}^k H_{CB}{}^{mnli} \right. \\
 &\left. +\frac{\rho-1}{2}\normalvector_A\normalvector_C\tetrad_B{}^{[m}\inducedmetric^{n][k}\inducedmetric^{l]i}+\normalvector_A \normalvector_B\tetrad_C{}^{[m}\inducedmetric^{n][k}\inducedmetric^{l]i} \right)+\inducedmetric^{im}\normalvector_A\tetrad^C{}_m\partial_j \momenta_C{}^j,
	\end{split}
	\end{align}
	which explicitly and implicitly (through the super-metric $H_{AB}{}^{ijkl}$) contains the NGR parameter $\rho$. However, the quantity
	\begin{align}
	\begin{split}
	-{}^{\shift}\dot{\momenta}_A{}^{i}{}_j&=-\delta_j^i\partial_k\momenta_A{}^k,
	\end{split}
	\end{align}
 is equivalent to the corresponding term in TEGR. The next term contains $\rho$ implicitly, again through the 
 super-metric $H_{AB}{}^{ijkl}$ and does not include ${}^{\mathcal{A}}\lambda_{ij}$:
	
	\begin{align}
	\begin{split}
	-{}^{\partial}\dot{\momenta}_A{}^{ij}&=\frac{2\lapse\sqrt{\inducedmetric}}{\kappa}H_{BA}{}^{kl[ij]}T^B{}_{kl}+2\shift^{[i}\momenta_A{}^{j]}+2{}^{\mathcal{V}}\lambda^{[i} \tetrad_A{}^{j]}.
	\end{split}
	\end{align}
	
	The term related to $\mathcal{B}_\mathcal{A}\neq 0$ contributes with a very long expression compared to TEGR, namely:
	
	\begin{align}
	\begin{split}
	-{}^{\mathcal{BA}}\dot{\momenta}_A{}^i&=\partial_k \left[ \lapse (\rho-1) \inducedmetric^{n[k}\momenta_B{}^{i]}\normalvector_A \tetrad^B{}_n-\frac{\lapse (\rho-1)^2 \sqrt{\inducedmetric}}{2\kappa}\inducedmetric^{n[i}\inducedmetric^{k]j}\normalvector_A \normalvector_B T^B{}_{nj} \right]\\
	&+\lapse \inducedmetric^{im}\eta_{CD}\normalvector_A \tetrad^D{}_m\inducedmetric^{kl}T^C{}_{kj}\left(\frac{\rho-1}{2} \momenta_B{}^j \tetrad^B{}_l+\frac{(\rho-1)^2}{4\kappa}\sqrt{\inducedmetric}\inducedmetric^{jn}\normalvector_B T^B{}_{ln} \right)\\
	&+\frac{\lapse \kappa}{2\sqrt{\inducedmetric}} \inducedmetric_{jk}\inducedmetric^{l[n}\momenta_B{}^{j]}\momenta_D{}^k \tetrad^B{}_l \tetrad_A{}^i\tetrad^D{}_n-\frac{\lapse (\rho-1)^2 \sqrt{\inducedmetric}}{8\kappa}\inducedmetric^{jl}\inducedmetric^{kn}\normalvector_B\normalvector_D T^B{}_{jk}T^D{}_{ln}\tetrad_A{}^i\\
	&+\lapse (\rho-1) \normalvector_B \momenta_C{}^j T^B{}_{lj}\tetrad^C{}_m\tetrad_A{}^{(l}\inducedmetric^{m)i}+\frac{\lapse\kappa}{2\sqrt{\inducedmetric}}\inducedmetric^{in}\inducedmetric_{jl}\momenta_B{}^j \momenta_C{}^l \tetrad^B{}_k \tetrad^C{}_n \tetrad_A{}^k\\
	&-\frac{\lapse\kappa}{2\sqrt{\inducedmetric}} \inducedmetric^{jk}\eta_{AD}\momenta_B{}^i\momenta_C{}^l\tetrad^B{}_j\tetrad^C{}_k\tetrad^D{}_l+\frac{2\lapse\sqrt{\inducedmetric}}{\kappa}\inducedmetric^{in}\inducedmetric^{l[k}\inducedmetric^{m]j}\eta_{AD}\normalvector_B \normalvector_D T^B{}_{jl}T^C{}_{mn}\tetrad^D{}_k \\
	&+ \frac{\lapse(\rho-1)}{2}\inducedmetric^{ik}\normalvector_B\momenta_A{}^j T^B{}_{kj}+\frac{\lapse \kappa}{\sqrt{\inducedmetric}}\momenta_B{}^{[i}\inducedmetric^{k]l}\inducedmetric_{jk}\momenta_A{}^j\tetrad^B{}_l.
	\end{split}
	\end{align}
	Thus, it is evident that computing the time evolution of constraints in NGR is considerably more cumbersome than in TEGR and $f(T)$ gravity, even when the Weitzenböck gauge is chosen. The next two parts of $\dot{\momenta}_A{}^i$ are the same as in TEGR and, in order to simplify the expression, can be combined as:
 \begin{align}
     \begin{split}
         {}^{\mathcal{BS}}\dot{\momenta}_A{}^i-\frac{1}{2}{}^{\mathcal{BT}}\dot{\momenta}_A{}^i&=\frac{\lapse\kappa}{2\sqrt{\inducedmetric}} \left(\momenta_C{}^{(i}\momenta_B{}^{l)}\tetrad^C{}_j\tetrad^B{}_k\eta_{AD}\inducedmetric^{jk}\tetrad^D{}_l-\momenta_C{}^m\momenta_B{}^l\tetrad^C{}_j\tetrad^B{}_k\tetrad_A{}^{(j}\inducedmetric^{k)i}\inducedmetric_{ml}\right. \\
 & \left. +2\momenta_A{}^j\momenta_B{}^{(k}\inducedmetric^{i)l}\inducedmetric_{jk}\tetrad^B{}_l   -\momenta_A{}^i\momenta_B{}^j\tetrad^B{}_j-\tetrad_A{}^i\momenta_C{}^m\momenta_B{}^{(l}\inducedmetric^{k)j}\tetrad^C{}_k\tetrad^B{}_j\inducedmetric_{ml} \right. \\
 & \left. +\frac{1}{2}\tetrad_A{}^i\momenta_C{}^k\momenta_B{}^j\tetrad^C{}_k\tetrad^B{}_j \right).
     \end{split}
 \end{align}
 
	Finally, there exists no part proportional to ${}^{\mathcal{A}}\lambda_{ij}$, while the term proportional to ${}^{\mathcal{V}}\lambda_i $ is the same as the corresponding one for TEGR:
	\begin{align}
 \label{eq:GaugeNGRLV}
	\begin{split}
	-{}^{\lambda\mathcal{V}}\dot{\momenta}_A{}^{ij}&=-\frac{\kappa}{\sqrt{\inducedmetric}}\inducedmetric^{im}\normalvector_A\tetrad^C{}_m\momenta_C{}^j-\frac{\kappa}{\sqrt{\inducedmetric}}\normalvector^B\momenta_B{}^j\tetrad_A{}^i\\
 &+T^B{}_{kj}\left(\tetrad_A{}^k \tetrad_B{}^i+\normalvector_A\normalvector_B \inducedmetric^{ik}\right)+2 T^B{}_{kl}\tetrad_B{}^k\tetrad_A{}^{(j}\inducedmetric^{l)i}.
	\end{split}
	\end{align}
	These are the Hamilton equations that, together with the primary constraints, can determine the evolution of the fields. The result can be used to confirm the findings in Ref. \cite{Cheng:1988zg} (which also assumes the Weitzenbock gauge) according to which the evolution of the primary constraints is generally not first class. To conclude, we point out that one can also consider calculations of Poisson brackets to simplify the above equations, by taking for instance the relations occurring in TEGR, which is surely more well studied. However, even doing so, it is clear that NGR calculations are still  very lengthy, if compared to TEGR.

	\section{Discussion and Conclusions}
	\label{sec:conc}
 In this paper, for the first time, we have presented the Hamiltonian (Eq. \eqref{eq:HfTNGR}) and the Hamilton equations in extensions of TEGR. In particular, we considered  $f(T)$ gravity and NGR, both in the covariant formulation \cite{Krssak:2015oua,Krssak:2018ywd} and in the Weitzenböck gauge. The Hamilton equations for $f(T_\mathrm{NGR})$-gravity are presented in App. \ref{sec:Heqscov} and, from this expression, the special cases of $f(T)$ gravity, NGR and TEGR can be recovered. As we expected, we found that the $f(T)$ Hamilton equations have a different structure than those of $f(R)$ gravity. Though we did not provide the explicit expression for the TEGR Hamilton equations, from the computations in Sec. \ref{sec:HeqfT}, it is straightforward to realize that our results are consistent with those in Ref. \cite{Pati:2022nwi}, except for a few typos. 

 In future work, these findings can be applied in two major directions. On the one hand, it is possible to use these results for the purposes of numerical relativity, as indicated in \cite{Capozziello:2021pcg,Pati:2022nwi}, where the case of TEGR was discussed. From this point of view, the present work could represent a first step in investigating numerical relativity for extended teleparallel theories of gravity. On the other hand, another direction is to investigate the evolution of the constraints. If they do not vanish on the primary constraint surface, they will be considered second class and must be enforced to vanish. This introduces more constraints, as discussed in Ref.~\cite{Blixt:2020ekl} (and references therein). Moreover, in Refs.~\cite{Blagojevic:2020dyq} and \cite{Tomonari:2024ybs,Tomonari:2024lpv} the explicit expressions of additional constraints, respectively in $f(T)$ and NGR, are considered. Also note that Hamilton's equations must be evaluated on the constraint surface in order to be equivalent to the Lagrangian equations of motion. In particular, the case of gauge fixing is interesting in both cases. First, not all observers $u^\mu=e_{\hat{0}}{}^\mu $\footnote{Here, we wrote $\hat{0}$ to emphasize that this is the temporal component of a Lorentz index.} admit global foliation \cite{Borowiec:2013kgx,Blixt:2024aej}, even in GR, although locally a foliation can always be made \cite{Baumgarte:2002jm}. In GR, this is not a big issue, since one can always perform a Lorentz transformation to obtain foliation in agreement with the field equations. In the covariant formulation of teleparallel gravity, any Lorentz transformation can still be applied, but a Lorentz transformed spin connection may be required to solve the field equations. Hence, the safest approach is to transform the tetrad and spin connection together, since this defines an equivalence class of solutions to the field equations. Thus, foliation can always be obtained in the covariant formulation. Similar considerations apply to TEGR in the Weitzenböck gauge, as a consequence of Lorentz invariance of type II (see \cite{Blixt:2022rpl} for the definition). In other words, a Lorentz transformed tetrad solution, fixing the spin connection, is again a solution. Since foliation has been assumed throughout this work, it would be interesting to look more deeply into the conditions for foliation, considering relevant spacetimes in $f(T)$ gravity and NGR as test bed. 

For numerical relativity, it is very important to choose a gauge to avoid instabilities. It is yet to be investigated whether the Weitzenböck gauge admits (strong) hyperbolicity. A gauge which has not yet been discussed in this work is the gauge fixing lapse function and shift vector, which is a valid choice since teleparallel theories are diffeomorphism-invariant. For example, the gauge $\lapse=1$, $\shift^i=0$ would further simplify the Hamiltonian and the Hamilton equations. However, such a gauge choice is known not to work well for numerical relativity in GR. This led to the development of the BSSN-formalism \cite{Baumgarte:2002jm} and we would expect something similar in teleperallel numerical relativity.

 It is often stressed that it is important to follow the covariant approach of teleparallel gravity \cite{Krssak:2015oua,Krssak:2018ywd} to restore Lorentz invariance of type I (following the definitions of \cite{Blixt:2022rpl}). It has been argued in \cite{Golovnev:2021omn,Blixt:2018znp,Blixt:2019mkt} that, working in the Weitzenböck gauge (opposed to the covariant formulation), the number of degrees of freedom is not affected. Furthermore, this is known to be true in the case of TEGR \cite{Blagojevic:2000qs}. This work can be used to give the first explicit proof of this statement by calculating the time evolution of the primary constraints associated with Lorentz invariance of type I defined by Eq. \eqref{eq:LConstraints}. If it turns out that the time evolution vanishes on the constraint surface, then the primary constraints are of first class and the number of degrees of freedom is indeed unaffected. Furthermore, the covariant formulation could have advantages in numerical relativity and when considering energy and entropy \cite{Gomes:2022vrc}.  
 
Furthermore, in teleparallel theories of gravity, calculating the evolution of constraints is very complicated \cite{Blixt:2020ekl}, since the requirement that the constraints are conserved in time appears to be very different, depending on the background \cite{Cheng:1988zg}. In the case of $f(T)$ gravity, the difficulty in calculating the time evolution of the constraints has led to some conflicting results regarding the number of degrees of freedom \cite{Li:2011rn,Blagojevic:2020dyq,Ferraro:2018tpu}. The results obtained in this article can be used to independently check which is the correct result. In the case of NGR, it is evident, from the Hamilton equations, that the calculation of the time evolution of constraints is indeed very lengthy as indicated in Ref. \cite{Cheng:1988zg}.

Quantities that are theory-dependent may also be simplified by selecting the given model and the related functional action. To this purpose, one can rely on the presence of symmetries, which aim to reduce the dynamics and allow to shorten the overall computations. Models containing symmetries can be selected by means of the so-called Noether Symmetry Approach \cite{Bajardi:2022ypn, Acunzo:2021gqc, Urban:2020lfk, Bajardi:2020xfj}, a selection criterion aimed at finding easily-handled theories and related conserved quantities. Clearly, one must check whether selected models are experimentally viable, by comparing the corresponding field equation solutions with  observations.

	\section*{Acknowledgments}

	This paper is based upon work from COST Action CA21136 {\it Addressing observational tensions in cosmology with systematics and fundamental physics} (CosmoVerse) supported by European Cooperation in Science and Technology.
	The authors acknowledge the Istituto Nazionale di Fisica Nucleare (INFN), Sezione di Napoli, \textit{iniziative specifiche} GINGER,  QGSKY, and MOONLIGHT2.
    S.C. thanks the  {\it Gruppo Nazionale di Fisica Matematica} of {\it Istituto Nazionale di Alta Matematica} for the support.

	\appendix
	
	\section{The Hamilton equations for covariant $f(T_\mathrm{NGR})$ gravity}
	\label{sec:Heqscov} 
	
		A very useful equation is the variation of the super-metric with respect to the tetrads. It generalizes the result found in \cite{Pati:2022nwi}. It yields:
	\begin{align}
	\begin{split}
	\frac{\delta H_{CB}{}^{mnkl}}{\delta \tetrad^A{}_i}&=\tetrad_A{}^m H_{CB}{}^{inlk}+\tetrad_A{}^n H_{CB}{}^{imkl}+\tetrad_A{}^k H_{CB}{}^{mnli}+\tetrad_A{}^l H_{CB}{}^{mnik}+c_2\normalvector_A\normalvector_C\tetrad_B{}^{[m}\inducedmetric^{n][k}\inducedmetric^{l]i}\\
 &+c_2\normalvector_A \normalvector_B \inducedmetric^{i[m}\inducedmetric^{n][k}\tetrad_C{}^{l]} +c_3\normalvector_A\normalvector_C\inducedmetric^{i[m}\inducedmetric^{n][k}\tetrad_B{}^{l]}+c_3\normalvector_A \normalvector_B\tetrad_C{}^{[m}\inducedmetric^{n][k}\inducedmetric^{l]i},
	\end{split}
	\end{align}
 where the super-metric $H_{AB}{}^{ijkl}$ has been defined in Eq. \eqref{eq:TorScalar}. Note that, after the variation, the above expression will always be contracted with $T^C{}_{mn}T^B{}_{kl}$, so that one can use symmetry properties to obtain
 \begin{align}
 \label{eq:Hsimple}
	\begin{split}
	\frac{\delta H_{CB}{}^{mnkl}}{\delta \tetrad^A{}_i}&\equiv 2\tetrad_A{}^m H_{CB}{}^{inlk}+2\tetrad_A{}^k H_{CB}{}^{mnli}+2c_2\normalvector_A\normalvector_C\tetrad_B{}^{[m}\inducedmetric^{n][k}\inducedmetric^{l]i}+2c_3\normalvector_A\normalvector_C\inducedmetric^{i[m}\inducedmetric^{n][k}\tetrad_B{}^{l]}.
	\end{split}
	\end{align}
	We again present the $f(T_\mathrm{NGR})$ Hamiltonian derived in Sec. \ref{sec:Ham} 
	\begin{align}
	\begin{split}
	\mathcal{H}_{f(T_\mathrm{NGR})}&=\lapse  \left[ \frac{\sqrt{\inducedmetric}\phi}{2\kappa} \mathcal{B}_\mathcal{V}{}^{\mathcal{V}}C^i {}^{\mathcal{V}} C_i-\frac{\sqrt{\inducedmetric}\phi}{2\kappa} \mathcal{B}_\mathcal{A}{}^{\mathcal{A}}C^{ij} {}^{\mathcal{A}} C_{ij}-\frac{\sqrt{\inducedmetric}\phi}{2\kappa} \mathcal{B}_\mathcal{S}{}^{\mathcal{S}}C^{ij} {}^{\mathcal{S}} C_{ij}-\frac{3\sqrt{\inducedmetric}\phi}{2\kappa} \mathcal{B}_\mathcal{T}{}^{\mathcal{T}}C {}^{\mathcal{T}} C\right. \\& \left.-\frac{\sqrt{\inducedmetric}}{2\kappa}{}^3 \mathbb{T}+\frac{\sqrt{\inducedmetric}V(\phi)}{2\kappa}-\normalvector^A\partial_i \momenta_A{}^i+\momenta_A{}^i\spinconnection^A{}_{Bi}\normalvector^B \right]  \\
	&+\shift^j\left[ -\tetrad^A{}_j \partial_i \momenta_A{}^i+\momenta_A{}^i\spinconnection^A{}_{Ci} \tetrad^C{}_j-\momenta_A{}^i T^A{}_{ij} \right] -{}^\alpha \lambda {}^\alpha \pi-{}^\beta \lambda_i {}^\beta \pi^i-{}^\phi \lambda {}^\phi \pi \\
	&-\lambda_{AB}\left( P^{[A}{}_D\eta^ {B]C}\Lorentz_C{}^D+\momenta_C{}^i\eta^{C[B}\tetrad^{A]}{}_i \right)-{}^\mathcal{S}\lambda_{ij}\frac{{}^\mathcal{S} \pi^{ij}\kappa}{\phi \sqrt{\gamma}}-{}^\mathcal{T}\lambda \frac{{}^\mathcal{T}\pi \kappa}{\phi \sqrt{\inducedmetric}}  \\
	&-{}^{\mathcal{V}}\lambda_i\left(\frac{{}^{\mathcal{V}} \pi^i\kappa}{\sqrt{\inducedmetric}}+c_3 T^B{}_{jk}\inducedmetric^{ik}\inducedmetric^{jl}\tetrad^A{}_l\eta_{AB}\right)-{}^{\mathcal{A}}\lambda_{ij}\left(\frac{{}^{\mathcal{A}} \pi^{ij}\kappa}{\sqrt{\inducedmetric}}+c_2  \inducedmetric^{ik}\inducedmetric^{jl}T^B{}_{kl}\normalvector_B \right).
	\end{split}
	\end{align}
	The Hamiltonian constraint is given by:
	\begin{align}
	\begin{split}
	-{}^\alpha \dot{\pi}=\frac{\delta H}{\delta \alpha}&=\frac{\sqrt{\inducedmetric}\phi}{2\kappa} \mathcal{B}_\mathcal{V}{}^{\mathcal{V}}C^i {}^{\mathcal{V}} C_i-\frac{\sqrt{\inducedmetric}\phi}{2\kappa} \mathcal{B}_\mathcal{A}{}^{\mathcal{A}}C^{ij} {}^{\mathcal{A}} C_{ij}-\frac{\sqrt{\inducedmetric}\phi}{2\kappa} \mathcal{B}_\mathcal{S}{}^{\mathcal{S}}C^{ij} {}^{\mathcal{S}} C_{ij}-\frac{3\sqrt{\inducedmetric}\phi}{2\kappa} \mathcal{B}_\mathcal{T}{}^{\mathcal{T}}C {}^{\mathcal{T}} C \\& -\frac{\sqrt{\inducedmetric}}{2\kappa}{}^3 \mathbb{T}+\frac{\sqrt{\inducedmetric}V(\phi)}{2\kappa}-\normalvector^A\partial_i \momenta_A{}^i+\momenta_A{}^i\spinconnection^A{}_{Bi}\normalvector^B,
	\end{split}
	\end{align}
	which is slightly simplified in the presence of primary constraints. The momenta constraint, on the other hand, is completely independent of the form of $f(T_\mathrm{NGR})$ theory and yields:
	\begin{align}
 \label{eq:momconstfTNGR}
	-{}^\beta \dot{\pi}_i=\frac{\delta H}{\delta \beta^i}=-\tetrad^A{}_i \partial_j \momenta_A{}^j+\momenta_A{}^j\spinconnection^A{}_{Cj} \tetrad^C{}_i-\momenta_A{}^j T^A{}_{ji} .
	\end{align}
	The evolution of the conjugate momenta with respect to the spatial tetrads is very lengthy, and for this reason it is presented here in the following form	
	\begin{align}
		\begin{split}
		-\dot{\momenta}_A{}^i&=\frac{\partial H}{\partial \tetrad^A{}_i}=-\lapse {}^{\lapse}\dot{\momenta}_A{}^i-\shift^j {}^{\shift}\dot{\momenta}_A{}^{i}{}_j-\partial_j {}^{\partial} \dot{\momenta}_A{}^{ij}-\lambda_{[BC]}{}^{\omega}\dot{\momenta}_A{}^{iBC}-\mathcal{B}_\mathcal{V}{}^{\mathcal{BV}}\dot{\momenta}_A{}^i-\mathcal{B}_\mathcal{A}{}^{\mathcal{BA}}\dot{\momenta}_A{}^i\\
		&-\mathcal{B}_\mathcal{S}{}^{\mathcal{BS}}\dot{\momenta}_A{}^i -\mathcal{B}_\mathcal{T}{}^{\mathcal{BT}}\dot{\momenta}_A{}^i  -{}^{\mathcal{V}}\lambda_j {}^{\lambda\mathcal{V}}\dot{\momenta}_A{}^{ij}-{}^{\mathcal{A}}\lambda_{[jk]} {}^{\lambda \mathcal{A}}\dot{\momenta}_A{}^{ijk}-{}^{\mathcal{S}}\lambda_{(jk)} {}^{\lambda \mathcal{S}}\dot{\momenta}_A{}^{ijk}-{}^{\mathcal{T}}\lambda {}^{\lambda \mathcal{T}}\dot{\momenta}_A{}^{i},
		\end{split}
	\end{align}
	where the explicit expressions of each term can be found in Eqs. \eqref{eq:fTNGRlapsemom}-\eqref{eq:fTNGRlT} below. Note that the expression depends on the specific theories which alter the number and character of the primary constraints. When considering a specific theory, it is guaranteed that the expression simplifies since the presence of primary constraints implies that the corresponding $\mathcal{B}$-term vanishes. On the other hand, if primary constraints are not present, then the corresponding Lagrange multiplier does not occur as well. In the covariant formulation, we also need to take into account the evolution of the conjugate momenta with respect to the Lorentz matrices. This expression is lengthy as well, so again we use a shortened notation
	\begin{align}
		\begin{split}
		-\dot{P}^A{}_B&=\frac{\partial H}{\partial \Lambda_A{}^B}=-\lapse {}^{\lapse}\dot{P}^A{}_B-\shift_i{}^{\shift}\dot{P}^A{}_B{}^i-\partial_i {}^{\partial}\dot{P}^A{}_B{}^i-\lambda_{[CD]}{}^{\omega}\dot{P}^A{}_B{}^{CD}-\mathcal{B}_\mathcal{V}{}^{\mathcal{BV}}\dot{P}^A{}_B-\mathcal{B}_\mathcal{A}{}^{\mathcal{BA}}\dot{P}^A{}_B\\
		&-{}^{\mathcal{V}}\lambda_i {}^{\lambda\mathcal{V}}\dot{P}^A{}_B{}^i-{}^{\mathcal{A}}\lambda_{[ij]} {}^{\lambda\mathcal{A}}\dot{P}^A{}_B{}^{ij},
		\end{split}
	\end{align}
 where the explicit expressions of the terms appearing in the above equation can be found in Eqs. \eqref{eq:fTNGRPlapse}-\eqref{eq:fTNGRPlA}. This part is only affected by the $\mathcal{V}$ and $\mathcal{A}$-part of the $\mathcal{VAST}$ decomposition. This is expected since Lorentz transformations are not associated with the symmetric part of the field equations. 
 
 The next term to evaluate is the time evolution of the momenta with respect to the scalar field, associated with the nonlinear extension of teleparallel theories

	\begin{align}
	\begin{split}
	-{}^\phi \dot{\pi}&=\frac{\delta H}{\delta \phi }=\lapse\left[ \mathcal{B}_\mathcal{V}\left(\frac{c_3^2 \sqrt{\inducedmetric} \inducedmetric^{jl}T^A{}_{ij}T^B{}_{kl}\tetrad_A{}^i \tetrad_B{}^k}{2\kappa} -\frac{\kappa \inducedmetric_{ij}\normalvector^A \normalvector^B \momenta_A{}^i \momenta_B{}^j}{2\sqrt{\inducedmetric} \phi^2}\right) \right.  \\
	& \left. +\mathcal{B}_\mathcal{A}\left(-\frac{c_2^2 \sqrt{\inducedmetric} \inducedmetric^{ik}\inducedmetric^{jl}\normalvector_A \normalvector_B T^A{}_{ij}T^B{}_{kl}}{2\kappa} +\frac{\kappa \inducedmetric^{kl}\inducedmetric_{i[j} \tetrad^A{}_{k]} \tetrad^B{}_l\momenta_A{}^i \momenta_B{}^j}{2\sqrt{\inducedmetric} \phi^2}\right) -\frac{\sqrt{\inducedmetric}}{2\kappa \phi}{}^3 \mathbb{T}  \right. \\
	& \left. +\frac{\mathcal{B}_\mathcal{S}\kappa }{2\sqrt{\inducedmetric} \phi^2}\left(\inducedmetric^{kl}\inducedmetric_{i(j} \tetrad^A{}_{k)} \tetrad^B{}_l\momenta_A{}^i \momenta_B{}^j  -\frac{\momenta_A{}^i\momenta_B{}^j\tetrad^A{}_i \tetrad^B{}_j }{3}\right) +\mathcal{B}_\mathcal{T}\frac{\kappa \momenta_A{}^i \momenta_B{}^j\tetrad^A{}_i \tetrad^B{}_j }{6\sqrt{\inducedmetric} \phi^2}+\frac{ \sqrt{\inducedmetric}}{2\kappa}\frac{\delta V(\phi)}{\delta \phi} \right]\\
	&-{}^\mathcal{V}\lambda_i \frac{\kappa \normalvector^A \momenta_A{}^i }{\sqrt{\inducedmetric} \phi^2}+{}^\mathcal{A}\lambda_{[ik]}\frac{\kappa \inducedmetric^{ij}\momenta_A{}^k \tetrad^A{}_j}{\sqrt{\inducedmetric} \phi^2}+{}^\mathcal{S}\lambda_{ik}\left(\frac{\kappa \inducedmetric^{j(k}\momenta_A{}^{i)}\tetrad^A{}_j}{\sqrt{\inducedmetric} \phi^2}-\frac{\kappa \inducedmetric^{jk}\momenta_A{}^l \tetrad^A{}_l \delta^i_j}{3\sqrt{\inducedmetric} \phi^2} \right)+{}^\mathcal{T}\lambda \frac{\kappa \momenta_A{}^i\tetrad^A{}_i }{3\sqrt{\inducedmetric} \phi^2}. 
	\end{split}
	\end{align}
The above quantity is very relevant for $f(T)$ gravity and for more details or discussion on the comparison with $f(R)$ gravity see Sec. \ref{sec:HeqfT}.

 Teleparallel theories of gravity are, like GR, invariant under diffeomorphism transformations and, as expected, the time evolution of lapse and shift are proportional to Lagrange multipliers:
	\begin{align}
	\dot{\alpha}=\frac{\delta H}{\delta {}^\alpha \pi}=-{}^\lapse \lambda, 
	\end{align}
	\begin{align}
	\dot{\beta}^i=\frac{\delta H}{\delta {}^\beta\pi_i}=-{}^\shift \lambda^i.
	\end{align}
 The time evolution of the spatial tetrad is dependent on the occurrence of primary constraints in the given theory. It reads:
	\begin{align}
	\begin{split}
	\dot{\theta}^A{}_i=\frac{\delta H}{\delta \pi_A{}^i}&= \lapse\left[\mathcal{B}_\mathcal{V}\left(\frac{\kappa \inducedmetric_{ij}\normalvector^A \normalvector^B \momenta_B{}^j}{\sqrt{\inducedmetric} \phi} -c_3 \inducedmetric^{jk}\eta_{BC}\normalvector^A T^B{}_{ji}\tetrad^C{}_k \right) \right. \\
	& \left. +\mathcal{B}_\mathcal{A}\left(-\frac{\kappa\inducedmetric^{kl}\inducedmetric_{i[j}\tetrad^A{}_{k]}\tetrad^B{}_l \momenta_B{}^j}{\sqrt{\inducedmetric} \phi }+c_2 \inducedmetric^{jk}\normalvector_B T^B{}_{ij}\tetrad^A{}_k \right)\right. \\
	&\left. -\mathcal{B}_\mathcal{S}\frac{ \kappa}{\sqrt{\inducedmetric} \phi }\left(\inducedmetric^{kl}\inducedmetric_{i(j}\tetrad^A{}_{k)}\tetrad^B{}_l \momenta_B{}^j -\frac{\momenta_B{}^j\tetrad^A{}_i \tetrad^B{}_j }{3}\right)-\mathcal{B}_\mathcal{T}\frac{\kappa \momenta_B{}^j\tetrad^A{}_i \tetrad^B{}_j}{3\sqrt{\inducedmetric} \phi}+\normalvector^B\omega^A{}_{Bi}+\partial_i\normalvector^A \right]\\
	&+\shift^j\left[\tetrad^B{}_j \omega^A{}_{Bi}-T^A{}_{ij}\right]+{}^\omega \lambda_{[CB]}\tetrad^B{}_i\eta^{AC}+{}^\mathcal{V}\lambda_i \frac{\kappa \normalvector^A}{\sqrt{\inducedmetric} \phi}+{}^\mathcal{A}\lambda_{[ik]}\frac{\kappa\inducedmetric^{jk}\tetrad^A{}_j}{\sqrt{\inducedmetric} \phi}\\
	&-{}^\mathcal{S}\lambda_{(kl)}\left(\frac{\kappa \delta^l_i \inducedmetric^{jk}\tetrad^A{}_j}{\sqrt{\inducedmetric}\phi} -\frac{\kappa\inducedmetric^{kl}\tetrad^A{}_i}{3\sqrt{\inducedmetric}\phi}\right)-{}^\mathcal{T}\lambda \frac{\kappa \tetrad^A{}_i}{3\sqrt{\inducedmetric}\phi}+\partial_i\left(\shift^j \tetrad^A{}_j \right).
	\end{split}
	\end{align}

The time evolution of the Lorentz matrices appears in the covariant formalism, as it is related to the primary constraints associated with the covariant formulation 
 
	\begin{align}
 \label{eq:LorentzEv}
	\begin{split}
	\dot{\Lambda}_A{}^B&=\frac{\delta H}{\delta P^A{}_B}= \lambda_{[DA]}\Lambda_C{}^B \eta^{CD}.
	\end{split}
	\end{align}
 Since the primary constraints are theory-independent, the same goes for the time evolution of the Lorentz matrices. Finally, the time evolution of the scalar field is determined by its associated Lagrange multiplier, which is very different from the case of $f(R)$ gravity (see Ref. \cite{Deruelle:2009zk}). It is:
	
	\begin{align}
	\dot{\phi}=\frac{\delta H}{\delta {}^\phi \pi}=-{}^\phi \lambda. 
	\end{align}
	In what follows, we present the explicit expression of all the terms which constitute $\dot{\momenta}_A{}^i$. Firstly, let us take into account ${}^{\lapse}\dot{\momenta}_A{}^i$, which yields:
	\begin{align}
 \label{eq:fTNGRlapsemom}
		\begin{split}
		 -{}^{\lapse}\dot{\momenta}_A{}^i&=-\frac{2\sqrt{\inducedmetric}}{\kappa}H_{CB}{}^{mn[ki]}T^C{}_{mn}\omega^B{}_{Ak}+\frac{\sqrt{\inducedmetric}}{2\kappa}\tetrad_A{}^i\left(V(\phi)-{}^3 \mathbb{T} \right) \\
		 &-\frac{\sqrt{\inducedmetric}}{\kappa}T^C{}_{mn}T^B{}_{kl}\left(\tetrad_A{}^m H_{CB}{}^{inlk}+\tetrad_A{}^k H_{CB}{}^{mnli}\right. \\
		 &\left. +c_2\normalvector_A\normalvector_C\tetrad_B{}^{[m}\inducedmetric^{n][k}\inducedmetric^{l]i}+c_3\normalvector_A \normalvector_B\tetrad_C{}^{[m}\inducedmetric^{n][k}\inducedmetric^{l]i} \right)+\inducedmetric^{im}\normalvector_A\tetrad^C{}_m\left(\partial_j \momenta_C{}^j-\momenta_B{}^j\omega^B{}_{Cj}\right).
		\end{split}
	\end{align}
 Notice that the above quantity depends on the specific theory considered through the super-metric $H_{AB}{}^{ijkl}$, which indeed is theory-dependent. On the other hand,
	
	\begin{align}
		\begin{split}
	 -{}^{\shift}\dot{\momenta}_A{}^{i}{}_j&= \momenta_B{}^{i}\omega^B{}_{Aj}-\delta_j^i\partial_k\momenta_A{}^k,
		\end{split}
	\end{align}
	takes the same form for all $f(T_\mathrm{NGR})$ theories. The term obtained with a derivative reads
		\begin{align}
	\begin{split}
	-{}^{\partial}\dot{\momenta}_A{}^{ij}&=\frac{2\lapse\sqrt{\inducedmetric}}{\kappa}H_{BA}{}^{kl[ij]}T^B{}_{kl}+2\shift^{[i}\momenta_A{}^{j]}+2{}^{\mathcal{V}}\lambda^{[i} \tetrad_A{}^{j]}+2{}^{\mathcal{A}}\lambda_{[kl]}c_2 \inducedmetric^{k[i}\inducedmetric^{j]l}\normalvector_A,
	\end{split}
	\end{align}
	and it is theory dependent as well. The term related to the primary constraints appearing in the covariant formulation is, as expected, theory independent and reads as:
	
		\begin{align}
	\begin{split}
	-{}^{\omega}\dot{\momenta}_A{}^{iBC}&=\delta^{[C}_A \eta^{B]D}\momenta_D{}^i.
	\end{split}
	\end{align}
	The next component of $\dot{\momenta}_A{}^i$ is:
		\begin{align}
	\begin{split}
	-{}^{\mathcal{BV}}\dot{\momenta}_A{}^i&=-2\lapse c_3 \inducedmetric^{j[k}\momenta_B{}^{i]}\eta_{CD}\normalvector^B \tetrad^D{}_j \omega^C{}_{Ak}+\frac{2\lapse c_3^2 \sqrt{\inducedmetric}\phi}{\kappa}\inducedmetric^{m[k}\inducedmetric^{i]l}\eta_{CD}T^B{}_{jl}\tetrad^D{}_m \tetrad_B{}^j  \omega^C{}_{Ak}\\
	&+\partial_k\left[2\lapse c_3\inducedmetric^{j[k}\momenta_B{}^{i]}\eta_{AD}\normalvector^B\tetrad^D{}_j+\frac{2\lapse c_3^2\phi\sqrt{\inducedmetric}}{\kappa}\inducedmetric^{jn}\inducedmetric^{m[i}\inducedmetric^{k]l}\eta_{BE}\eta_{AD}T^B{}_{jl}\tetrad^D{}_m \tetrad^E{}_n  \right]\\
	&+\lapse c_3\inducedmetric^{im}\normalvector_A\tetrad^C{}_m\momenta_C{}^j T^B{}_{kj}\tetrad_B{}^k-\frac{\lapse\kappa}{\sqrt{\inducedmetric}\phi} \inducedmetric^{im}\normalvector_A\tetrad^C{}_m\inducedmetric_{jk}\normalvector^B\momenta_B{}^k\momenta_C{}^j\\
	&+\frac{\lapse}{2\sqrt{\inducedmetric}\phi}\inducedmetric_{jk}\normalvector^B\normalvector^D\momenta_B{}^j\momenta_D{}^k\tetrad_A{}^i+\frac{\lapse c_3^2\sqrt{\inducedmetric}\phi}{2\kappa}\inducedmetric^{jm}\inducedmetric^{kn}T^B{}_{jk}T^D{}_{ln}\tetrad_A{}^i\tetrad^E{}_m\tetrad_D{}^l\\
	&+2\lapse\inducedmetric^{i(m}\inducedmetric^{l)k}\eta_{AD}\eta_{CE}\normalvector^B \momenta_B{}^j T^C{}_{lj}\tetrad^D{}_k\tetrad^E{}_m+\frac{\lapse\kappa}{\sqrt{\inducedmetric}\phi} \eta_{AD}\normalvector^B\normalvector^C\momenta_B{}^k\momenta_C{}^i\tetrad^D{}_k\\
	&+\frac{2\lapse c_3^2 \sqrt{\inducedmetric}\phi}{\kappa}\inducedmetric^{ij}T^B{}_{jl}T^C{}_{mn}\tetrad_A{}^{[n} \tetrad_B{}^{l]} \tetrad_C{}^m  -\frac{\lapse c_3^2\sqrt{\inducedmetric}}{\kappa}\inducedmetric^{ln}T^B{}_{jl}T^C{}_{mn}\tetrad_A{}^j \tetrad_B{}^i\tetrad_C{}^m\\
	&+\lapse c_3\inducedmetric^{ik}\eta_{AC}\normalvector^B\momenta_B{}^jT^C{}_{jk}+\frac{\lapse c_3^2 \sqrt{\inducedmetric}\phi}{\kappa}\inducedmetric^{ij}\inducedmetric^{km}\inducedmetric^{ln}\eta_{AB}\eta_{CD}T^B{}_{jk}T^C{}_{lm}\tetrad^D{}_n, 
	\end{split}
	\end{align}
	which only appears for $\mathcal{B}_\mathcal{V}\neq 0$, and has regained interest recently \cite{Bahamonde:2024zkb}. In TEGR and $f(T)$ gravity, $\mathcal{B}_\mathcal{A}=0$, while it is not-vanishing for NGR. Hence, the next expression is worth to be investigated in the case of NGR (see Sec. \ref{sec:HeqNGR} for more details and discussion) 
		\begin{align}
	\begin{split}
	-{}^{\mathcal{BA}}\dot{\momenta}_A{}^i&=-2 \lapse c_2 \inducedmetric^{n[i}\momenta_B{}^{k]}\normalvector_C \tetrad^B{}_n\omega^C{}_{Ak}-\frac{2\lapse c_2^2 \sqrt{\inducedmetric}\phi}{\kappa}\inducedmetric^{n[k}\inducedmetric^{i]j}\normalvector_C \normalvector_B T^B{}_{nj}\omega^C{}_{Ak}\\
	&+\partial_k \left[-2 \lapse c_2 \inducedmetric^{n[k}\momenta_B{}^{i]}\normalvector_A \tetrad^B{}_n-\frac{2\lapse c_2^2 \sqrt{\inducedmetric}\phi}{\kappa}\inducedmetric^{n[i}\inducedmetric^{k]j}\normalvector_A \normalvector_B T^B{}_{nj} \right]\\
	&+\lapse \inducedmetric^{im}\eta_{CD}\normalvector_A \tetrad^D{}_m\inducedmetric^{kl}T^C{}_{kj}\left(-c_2 \momenta_B{}^j \tetrad^B{}_l+\frac{c_2^2\phi}{\kappa}\sqrt{\inducedmetric}\inducedmetric^{jn}\normalvector_B T^B{}_{ln} \right)\\
	&+\frac{\lapse \kappa}{2\sqrt{\inducedmetric}\phi} \inducedmetric_{jk}\inducedmetric^{l[n}\momenta_B{}^{j]}\momenta_D{}^k \tetrad^B{}_l \tetrad_A{}^i\tetrad^D{}_n-\frac{\lapse c_2^2 \sqrt{\inducedmetric}\phi}{2\kappa}\inducedmetric^{im}\inducedmetric^{jl}\inducedmetric^{kn}\eta_{AC}\normalvector_B\normalvector_D T^B{}_{jk}T^D{}_{ln}\tetrad^C{}_m\\
	&-2\lapse c_2 \normalvector_B \momenta_C{}^j T^B{}_{lj}\tetrad^C{}_m\tetrad_A{}^{(l}\inducedmetric^{m)i}+\frac{\lapse\kappa}{2\sqrt{\inducedmetric}\phi}\inducedmetric^{in}\inducedmetric_{jl}\momenta_B{}^j \momenta_C{}^l \tetrad^B{}_k \tetrad^C{}_n \tetrad_A{}^k\\
	&-\frac{\lapse\kappa}{2\sqrt{\inducedmetric}\phi} \inducedmetric^{jk}\eta_{AD}\momenta_B{}^i\momenta_C{}^l\tetrad^B{}_j\tetrad^C{}_k\tetrad^D{}_l+\frac{2\lapse\sqrt{\inducedmetric}\phi}{\kappa}\inducedmetric^{in}\inducedmetric^{l[k}\inducedmetric^{m]j}\eta_{AD}\normalvector_B \normalvector_D T^B{}_{jl}T^C{}_{mn}\tetrad^D{}_k \\
	&-\lapse c_2\inducedmetric^{ik}\normalvector_B\momenta_A{}^j T^B{}_{kj}+\frac{\lapse \kappa}{\sqrt{\inducedmetric}\phi}\momenta_B{}^{[i}\inducedmetric^{k]l}\inducedmetric_{jk}\momenta_A{}^j\tetrad^B{}_l.
	\end{split}
	\end{align}
	It is clear that this expression is very lengthy, which makes NGR a quite cumbersome theory to study. The next two expressions are necessary for the propagation of a massless spin-2 field
 \begin{align}
     \begin{split}
         -{}^{\mathcal{BS}}\dot{\momenta}_A{}^i&=\frac{\lapse\kappa}{2\sqrt{\inducedmetric}\phi}\left(\momenta_C{}^m\momenta_B{}^l\tetrad^C{}_j\tetrad^B{}_k\tetrad_A{}^{(j}\inducedmetric^{k)i}\inducedmetric_{ml}-\momenta_C{}^{(i}\momenta_B{}^{l)}\tetrad^C{}_j\tetrad^B{}_k \eta_{AD}\inducedmetric^{jk}\tetrad^D{}_l \right)\\
         &-\frac{\lapse\kappa}{2\sqrt{\inducedmetric}\phi}\left(\momenta_A{}^j\momenta_B{}^k\tetrad^B{}_l\inducedmetric^{il}\inducedmetric_{jk}+\momenta_A{}^j\momenta_B{}^i\tetrad^B{}_j-\frac{2}{3}\momenta_A{}^i \momenta_B{}^j\tetrad^B{}_j\right)\\
         &+\frac{\lapse\kappa}{2\sqrt{\inducedmetric}\phi}\tetrad_A{}^i\left(\momenta_C{}^m\momenta_B{}^{(l}\inducedmetric^{k)j}\tetrad^C{}_k\tetrad^B{}_j\inducedmetric_{ml}-\frac{1}{3}\momenta_C{}^k\momenta_B{}^j\tetrad^C{}_k\tetrad^B{}_j\right),
     \end{split}
 \end{align}
	and
		\begin{align}
	\begin{split}
	-{}^{\mathcal{BT}}\dot{\momenta}_A{}^i   &=\frac{\lapse\kappa}{6\sqrt{\inducedmetric}\phi} \tetrad_A{}^i\momenta_B{}^j\momenta_D{}^k\tetrad^B{}_j\tetrad^D{}_k-\frac{\lapse\kappa}{3\sqrt{\inducedmetric}\phi} \momenta_A{}^i\momenta_B{}^j\tetrad^B{}_j.
	\end{split}
	\end{align}
These two terms can be combined to simplify the final expression according to Eq. \eqref{eq:momsquared}. To obtain the correct gravitational behavior one needs to require $c_3=1$. If we require the theory to be ghost-free ($\mathcal{B}_\mathcal{V}=0$) and further require $\mathcal{B}_\mathcal{S}$ ($\mathcal{B}_\mathcal{T}$) to be zero, then $\mathcal{B}_\mathcal{T}$ ($\mathcal{B}_\mathcal{S}$) will be vanishing as well. Such a theory, hence, yields $\mathcal{B}_\mathcal{V}=\mathcal{B}_\mathcal{S}=\mathcal{B}_\mathcal{T}=0$ and the only propagating field is a Kalb-Ramond (pseudo-vector) field occurring at linear order of perturbation \cite{BeltranJimenez:2019nns}. The next term, required to avoid ghosts, is

		\begin{align}
	\begin{split}
	 -{}^{\lambda\mathcal{V}}\dot{\momenta}_A{}^{ij}&=-2\inducedmetric^{j[i}\tetrad_B{}^{k]}\omega^B{}_{Ak}-\frac{\kappa}{\sqrt{\inducedmetric}\phi}\inducedmetric^{im}\normalvector_A\tetrad^C{}_m\momenta_C{}^j\\
	 &-\frac{\kappa}{\sqrt{\inducedmetric}\phi}\normalvector^B\momenta_B{}^j\tetrad_A{}^i+c_3T^B{}_{kj}\left(\tetrad_A{}^k \tetrad_B{}^i+\normalvector_A\normalvector_B \inducedmetric^{ik}\right)+2 c_3T^B{}_{kl}\tetrad_B{}^k\tetrad_A{}^{(j}\inducedmetric^{l)i}.
	\end{split}
	\end{align}
	and occurs in TEGR, $f(T)$ gravity and NGR, unlike the following one that is absent in NGR
		\begin{align}
	\begin{split}
	 -{}^{\lambda \mathcal{A}}\dot{\momenta}_A{}^{ijk}&=-2c_2\inducedmetric^{kl}\inducedmetric^{ij}\normalvector_B\omega^B{}_{Al}-c_2\normalvector_A\tetrad_B{}^i\inducedmetric^{jn}\inducedmetric^{km}T^B{}_{nm}+\frac{\kappa}{\sqrt{\inducedmetric}\phi}\inducedmetric^{jl}\momenta_B{}^k\tetrad^B{}_l\tetrad_A{}^i\\
	 &+2c_2 \inducedmetric^{i[k}\inducedmetric^{j]m}\normalvector_B T^B{}_{ml}\tetrad_A{}^l+2c_2 \inducedmetric^{im}\normalvector_B T^B{}_{nm}\tetrad_A{}^{[k}\inducedmetric{}^{j]n}+\frac{2\kappa}{\sqrt{\inducedmetric}\phi} \momenta_B{}^k \tetrad^B{}_l\tetrad_A{}^{[l}\inducedmetric^{j]i}+\frac{\kappa}{\sqrt{\inducedmetric}\phi}\inducedmetric^{ik}\momenta_A{}^j.
	\end{split}
	\end{align}
	The last two terms, namely
		\begin{align}
	\begin{split}
	-{}^{\lambda \mathcal{S}}\dot{\momenta}_A{}^{ijk}&=\frac{1}{\sqrt{\inducedmetric}\phi}\tetrad_A{}^i\left(\inducedmetric^{kl}\momenta_B{}^{j}\tetrad^B{}_l-\frac{1}{3}\inducedmetric^{jk}\momenta_ B{}^l\tetrad^B{}_l \right)-\frac{\kappa}{\sqrt{\inducedmetric}\phi}\left(\inducedmetric^{ik}\momenta_A{}^j-\frac{1}{3}\inducedmetric^{jk}\momenta_A{}^i \right)\\
	&+\frac{2\kappa}{\sqrt{\inducedmetric}\phi}\eta_{AD}\tetrad^D{}_l\tetrad^B{}_m\left(\inducedmetric^{i(k}\inducedmetric^{l)m}\momenta_B{}^j-\frac{1}{3}\inducedmetric^{ik}\inducedmetric^{jl}\momenta_B{}^m \right),
	\end{split}
	\end{align}
	and
		\begin{align}
   \label{eq:fTNGRlT}
	\begin{split}
	 -{}^{\lambda \mathcal{T}}\dot{\momenta}_A{}^{i}&=\frac{\kappa}{3\sqrt{\inducedmetric}\phi}\left(\momenta_B{}^j\tetrad^B{}_j\tetrad_A{}^i-\momenta_A{}^i\right),
	\end{split}
	\end{align}
 cannot appear if we require  a propagating spin-2 field. Therefore, the above quantities are not present in the case of viable models within TEGR, $f(T)$ gravity or NGR. 

 We now present the explicit expressions for the time evolution of the momenta constraints. The first one is implicitly theory-dependent through the super-metric $H_{AB}{}^{ijkl}$:
	
	\begin{align}
	\label{eq:fTNGRPlapse}
  \begin{split}
		- {}^{\lapse}\dot{P}^A{}_B&=- \left( \Lambda^{-1}\right)^A{}_D\omega^C{}_{Bi}\normalvector^D\momenta_C{}^i -(H_{CE}{}^{kijl}-H_{CE}{}^{iljk})\frac{ \sqrt{\inducedmetric} \left(\Lambda^{-1} \right)^A{}_D \omega^C{}_{Bi}  T^E{}_{jl}\tetrad^D{}_k}{\kappa}.
		\end{split}
	\end{align}
	The following one, namely ${}^{\shift}\dot{P}^A{}_B{}^i$, reads:
		\begin{align}
	\begin{split}
	-{}^{\shift}\dot{P}^A{}_B{}^i&= -\momenta_C{}^{j} \left(\Lambda^{-1} \right)^A{}_D \omega^C{}_{Bi} \tetrad^D{}_j,
	\end{split}
	\end{align}
 and, as expected due to the close relation to the shift vector, is theory independent. The following quantity depends explicitly on the given theory because of the presence of $c_2$ and $c_3$:
	
		\begin{align}
	\begin{split}
	- {}^{\partial}\dot{P}^A{}_B{}^i&= \left[\lapse \left(\Lambda^{-1} \right)^A{}_D \normalvector^D \pi_B{}^i-\shift^i \left(\Lambda^{-1} \right)^A{}_D \momenta_B{}^k \right. \\
	& \left. +\frac{\lapse}{\kappa}\left(H_{BC}{}^{kijl}-H_{BC}{}^{iljk} \right)\sqrt{\inducedmetric} \left(\Lambda^{-1} \right)^A{}_D T^C{}_{jl} \tetrad^D{}_k \right. \\
	& \left. +2{}^\mathcal{A}\lambda_{[jk]}c_2 \left(\Lambda^{-1} \right)^A{}_D \inducedmetric^{ik}\inducedmetric^{jl} \normalvector_B \tetrad^D{}_l +2{}^\mathcal{V}\lambda_l c_3 \left(\Lambda^{-1} \right)^A{}_D \inducedmetric^{i[j}\inducedmetric^{l]k}\eta_{BC} \tetrad^C{}_j\tetrad^D{}_k \right],
	\end{split}
	\end{align}
 and also implicitly through the super metric $H_{AB}{}^{ijkl}$. A very simple expression comes from the part related to the primary constraints associated with the covariant formulation, that is
	
		\begin{align}
	\begin{split}
	-{}^{\omega}\dot{P}^A{}_B{}^{CD}&=\eta^{A[C}P^{D]}{}_B.
	\end{split}
	\end{align}
	The above term is obviously theory independent, as related to the primary constraints. The next part of the general momentum is included in the type of NGR theory that recently regained interest \cite{Bahamonde:2024zkb}
		\begin{align}
	\begin{split}
	-{}^{\mathcal{BV}}\dot{P}^A{}_B&=-2\lapse c_3 \left(\Lambda^{-1} \right)^A{}_D \omega^C{}_{Bi} \tetrad_C{}^{[k}\momenta_E{}^{i]}\normalvector^E  \tetrad^D{}_k  \\
	&-\frac{2\lapse c_3^2 \sqrt{\inducedmetric}\phi}{\kappa}  \left(\Lambda^{-1} \right)^A{}_D \omega^C{}_{Bi} \eta^{BD} T^E{}_{jk}\tetrad_B{}^{[k}\tetrad_C{}^{i]}\tetrad_E{}^j \\
	&+\partial_i \left[2\lapse c_3 \left(\Lambda^{-1} \right)^A{}_D \tetrad_B{}^{[k}\momenta_C{}^{i]}\normalvector^C \tetrad^D{}_k   +\frac{2\lapse c_3^2 \sqrt{\inducedmetric}\phi}{\kappa} \left(\Lambda^{-1} \right)^A{}_D \tetrad_B{}^{[i}\inducedmetric^{l]k}T^C{}_{jl}\tetrad^D{}_k  \tetrad_C{}^j  \right].
	\end{split}
	\end{align}
	The following term is of interest since it appears in NGR (see Sec. \ref{sec:HeqNGR})
		\begin{align}
	\begin{split}
	-{}^{\mathcal{BA}}\dot{P}^A{}_B&=2\lapse c_2 \left(\Lambda^{-1} \right)^A{}_D \omega^C{}_{Bi} \inducedmetric^{j[k}\momenta_E{}^{i]}\normalvector_C  \tetrad^D{}_k\tetrad^E{}_j\\
	&-\frac{2\lapse c_2^2 \sqrt{\inducedmetric}\phi}{\kappa}  \left(\Lambda^{-1} \right)^A{}_D \omega^C{}_{Bi} \inducedmetric^{i[l}\inducedmetric^{j]k}\normalvector_C \normalvector_E  T^E{}_{jl}\tetrad^D{}_k \\
	&+\partial_i \left[-2\lapse c_2 \left(\Lambda^{-1} \right)^A{}_D \inducedmetric^{j[k}\momenta_C{}^{i]} \normalvector_B \tetrad^C{}_j \tetrad^D{}_k+\frac{2\lapse c_2^2 \sqrt{\inducedmetric}\phi}{\kappa} \left(\Lambda^{-1} \right)^A{}_D \inducedmetric^{i[l}\inducedmetric^{j]k}\normalvector_B\normalvector_C T^C{}_{jl}\tetrad^D{}_k   \right]
	\end{split}
	\end{align}
	and it is absent in both TEGR and $f(T)$ gravity, where $\mathcal{B}_\mathcal{A}=0$. To avoid the propagation of ghosts, also the term below must be included in the total evaluation:
		\begin{align}
	\begin{split}
	- {}^{\lambda\mathcal{V}}\dot{P}^A{}_B{}^i&=2 c_3 \left(\Lambda^{-1} \right)^A{}_D \omega^C{}_{Bl} \tetrad_C{}^{[i}\inducedmetric^{l]k}  \tetrad^D{}_k.
	\end{split}
	\end{align}
	Finally, the last term listed here appears in TEGR and $f(T)$ gravity, but not in NGR: 
		\begin{align}
  \label{eq:fTNGRPlA}
	\begin{split}
	- {}^{\lambda\mathcal{A}}\dot{P}^A{}_B{}^{ij}&=2c_2 \left(\Lambda^{-1} \right)^A{}_D \omega^C{}_{Bk} \inducedmetric^{k[j}\inducedmetric^{i]l}\normalvector_C  \tetrad^D{}_l. 
	\end{split}
	\end{align}
This completes the expression for the Hamilton equations in $f(T_\mathrm{NGR})$ and, from this, it is straightforward to obtain the special cases including TEGR, $f(T)$ gravity and NGR. Furthermore, they appear without any gauge fixing. It is not difficult to get the Weitzenböck gauge result, or also to the gauge $\lapse=1$ and $\shift^i=0$, which further simplifies the result. 
 
	We have found, in Sec. \ref{sec:HeqfT}, that the expression for $\dot{\momenta}_A{}^i$ is mostly consistent with the TEGR expression found in \cite{Pati:2022nwi}, except for some typos. Since this case is of special importance, here we present the final corrected form of $\dot{\momenta}_A{}^i$, that is:
 	\begin{align}
  \label{eq:TEGRmom}
	\begin{split}
	-\dot{\momenta}_A{}^i&=\frac{\delta H}{\delta \tetrad^A{}_i}=\lapse \left(-\frac{2\sqrt{\inducedmetric}}{\kappa}H_{CB}{}^{mn[ki]}T^C{}_{mn}\omega^B{}_{Ak}-\frac{\sqrt{\inducedmetric}}{2\kappa}\tetrad_A{}^i{}^3 \mathbb{T}  \right. \\
	& \left. -\frac{\sqrt{\inducedmetric}}{\kappa}T^C{}_{mn}T^B{}_{kl}\left(\tetrad_A{}^m H_{CB}{}^{inlk}+\tetrad_A{}^k H_{CB}{}^{mnli}\right. \right. \\
	&\left. \left. -\frac{1}{2}\normalvector_A\normalvector_C\tetrad_B{}^{[m}\inducedmetric^{n][k}\inducedmetric^{l]i}+\normalvector_A \normalvector_B\tetrad_C{}^{[m}\inducedmetric^{n][k}\inducedmetric^{l]i} \right)+\inducedmetric^{im}\normalvector_A\tetrad^C{}_m\left(\partial_j \momenta_C{}^j-\momenta_B{}^j\omega^B{}_{Cj}\right) \right) \\
	& +\shift^j\left( \momenta_B{}^{i}\omega^B{}_{Aj}-\delta_j^i\partial_k\momenta_A{}^k \right)+\lambda_{[BA]}\eta^{BC}\momenta_C{}^i\\
	&+\partial_j \left(\frac{2\lapse\sqrt{\inducedmetric}}{\kappa}H_{BA}{}^{kl[ij]}T^B{}_{kl}+2\shift^{[i}\momenta_A{}^{j]}+2{}^{\mathcal{V}}\lambda^{[i} \tetrad_A{}^{j]}+{}^{\mathcal{A}}\lambda^{[ij]} \normalvector_A \right)\\
	&+\frac{\lapse\kappa}{2\sqrt{\inducedmetric}} \left(\momenta_C{}^{(i}\momenta_B{}^{l)}\tetrad^C{}_j\tetrad^B{}_k\eta_{AD}\inducedmetric^{jk}\tetrad^D{}_l-\momenta_C{}^m\momenta_B{}^l\tetrad^C{}_j\tetrad^B{}_k\tetrad_A{}^{(j}\inducedmetric^{k)i}\inducedmetric_{ml}+2\momenta_A{}^j\momenta_B{}^{(k}\inducedmetric^{i)l}\inducedmetric_{jk}\tetrad^B{}_l  \right. \\
 &\left. -\momenta_A{}^i\momenta_B{}^j\tetrad^B{}_j-\tetrad_A{}^i\momenta_C{}^m\momenta_B{}^{(l}\inducedmetric^{k)j}\tetrad^C{}_k\tetrad^B{}_j\inducedmetric_{ml}+\frac{1}{2}\tetrad_A{}^i\momenta_C{}^k\momenta_B{}^j\tetrad^C{}_k\tetrad^B{}_j\right)   \\
	&+{}^{\mathcal{V}}\lambda_j \left(-2\inducedmetric^{j[i}\tetrad_B{}^{k]}\omega^B{}_{Ak}-\frac{\kappa}{\sqrt{\inducedmetric}}\inducedmetric^{im}\normalvector_A\tetrad^C{}_m\momenta_C{}^j-  \inducedmetric^{ik}\inducedmetric^{jl}\eta_{AB}T^B{}_{kl} \right. \\
	& \left. -\frac{\kappa}{\sqrt{\inducedmetric}}\normalvector^B\momenta_B{}^j\tetrad_A{}^i+2 T^B{}_{kl}\tetrad_B{}^{[i}\inducedmetric^{j]k} \tetrad_A{}^l+2  \inducedmetric^{il}\eta_{BC}T^B{}_{kl}\tetrad^C{}_m\tetrad_A{}^{[m}\inducedmetric^{j]k}
	 \right)\\
	 &+{}^{\mathcal{A}}\lambda_{[jk]} \left(\inducedmetric^{kl}\inducedmetric^{ij}\normalvector_B\omega^B{}_{Al}+\frac{1}{2}\normalvector_A\tetrad_B{}^i\inducedmetric^{jn}\inducedmetric^{km}T^B{}_{nm}+\frac{\kappa}{\sqrt{\inducedmetric}}\inducedmetric^{jl}\momenta_B{}^k\tetrad^B{}_l\tetrad_A{}^i \right. \\
	 &\left. - \inducedmetric^{i[k}\inducedmetric^{j]m}\normalvector_B T^B{}_{ml}\tetrad_A{}^l- \inducedmetric^{im}\normalvector_B T^B{}_{nm}\tetrad_A{}^{[k}\inducedmetric{}^{j]n}+\frac{2\kappa}{\sqrt{\inducedmetric}} \momenta_B{}^k \tetrad^B{}_l\tetrad_A{}^{[l}\inducedmetric^{j]i}+\frac{\kappa}{\sqrt{\inducedmetric}}\inducedmetric^{ik}\momenta_A{}^j \right).
	\end{split}
	\end{align}
    This concludes the derivation.

	\bibliographystyle{ieeetr}
	\bibliography{references.bib}

\end{document}